%% file: red-blue-game.tex
\documentclass[reprint,pre,twoside,superscriptaddress,floatfix]{revtex4-1}

\input{red-blue-game.settings}

\begin{document}


\title{Noncooperative dynamics in election interference}

%%%%%%%%%%%%%%%%%%%%%%%%%%%%%%%%%%%%%%%%%%%%%%
%%                                          %%
%% Enter the authors here                   %%
%%                                          %%
%% Specify information, if available,       %%
%% in the form:                             %%
%%   <key>={<id1>,<id2>}                    %%
%%   <key>=                                 %%
%% Comment or delete the keys which are     %%
%% not used. Repeat \author command as much %%
%% as required.                             %%
%%                                          %%
%%%%%%%%%%%%%%%%%%%%%%%%%%%%%%%%%%%%%%%%%%%%%%

\author{David Rushing Dewhurst \thanks{david.dewhurst@uvm.edu}}
\affiliation{
MassMutual Center of Excellence in Complex Systems and Data Science,
Computational Story Lab,
Vermont Complex Systems Center, 
and Department of Mathematics and Statistics,
University of Vermont, Burlington, VT 05405}

\author{Christopher M. Danforth \thanks{chris.danforth@uvm.edu}}
\affiliation{
MassMutual Center of Excellence in Complex Systems and Data Science,
Computational Story Lab,
Vermont Complex Systems Center, 
and Department of Mathematics and Statistics,
University of Vermont, Burlington, VT 05405}

\author{Peter Sheridan Dodds \thanks{peter.dodds@uvm.edu}}
\affiliation{
MassMutual Center of Excellence in Complex Systems and Data Science,
Computational Story Lab,
Vermont Complex Systems Center, 
and Department of Mathematics and Statistics,
University of Vermont, Burlington, VT 05405}

\begin{abstract} % abstract
Foreign power interference in domestic elections is an existential threat to societies.
Manifested through myriad methods from war to words, such interference 
is a timely example of 
strategic interaction between economic and political agents. 
We model this interaction between rational game players 
as a continuous-time differential game, constructing 
an analytical model of this competition with a variety of payoff structures. 
All-or-nothing attitudes by only one player regarding the outcome of the game
lead to an arms race in which both countries spend increasing amounts on interference 
and counter-interference operations.
We then confront our model with data pertaining to the Russian interference in the 2016 United States
presidential election contest.
We introduce and estimate
a Bayesian structural time series model of election polls and social media posts by Russian Twitter
troll accounts. 
Our analytical model, while purposefully abstract and simple,
adequately captures many temporal characteristics of the election and social media activity.
We close with a discussion of our model's shortcomings and suggestions for future research.
\end{abstract}

\maketitle
\section{Introduction}\label{sec:intro}
In democratic and nominally-democratic countries, elections are societally and politically crucial 
events in which power is allocated \cite{renaud1987importance}.
In fully-democratic countries elections are the method of legitimate governmental
change \cite{elklit1997makes}.
One country, labeled ``Red'', wishes to influence the 
outcome of an election in another country, labeled ``Blue', because of the impact 
that elections in Blue have on Red's national interest.
Such attacks on democracies are not new.
It is estimated that the United States (U.S.) and Russia (and its
predecessor, 
the Soviet Union) often interfere in the elections of other nations and have consistently done this
since 1946 \cite{levin2016great}.
Though academic study of this area has increased \cite{shulman2012legitimacy},
we are unaware of any formal modeling of noncooperative dynamics in an election interference game.
Recent approaches to the study of this phenomenon have focused mainly on the compilation of 
coarse-grained (e.g., yearly frequency) panels of election interference events and qualitative analysis of 
this data
\cite{corstange2012taking,levin2019partisan}, and 
data-driven studies of the aftereffects and second-order effects of interference operations 
\cite{borghard2018confidence,levin2018voting}.
Attempts to create theoretical models of interference operations are less common.
These attempts include 
qualitative causal models of cyberoperation influence on voter preferences \cite{hansen2019doxing}
and models of the underlying reasons that a state may wish to interfere in the elections of another 
\cite{bubeck2019states}.
\medskip

\noindent
We
consider a two-player game in which one country wants to influence a two-candidate, zero-sum 
election taking place in another country. 
We think of Red
as the foriegn intelligence service of the influencing country and Blue as the domestic intelligence service 
of the country in which the election is held.
Red wants a particular candidate, which we will set to be candidate A without loss of generality,
to win the election, while Blue wants the effect of Red's interference to be 
minimized.
We derive a noncooperative, non-zero-sum differential game to describe this problem, 
and then explore the game numerically. 
We find that all-or-nothing attitudes by either Red or Blue can lead to arms-race conditions in
interference operations. 
In the event that one party credibly commits to playing a fixed, deterministic strategy, we derive further analytical 
results.
\medskip

\noindent
We then confront our model with data pertaining
to the 2016 U.S.\ presidential election contest, in which Russia interfered \cite{nyt2019}.
We fit 
a Bayesian structural time series model to election polls and social media posts authored by
Russian military intelligence-associated troll accounts.
We demonstrate that our model, though simple, captures many of the observed and inferred parameters'
dynamics.
We close by proposing some theoretical and empirical extensions to our work.
\section{Theory}\label{sec:theory}

\subsection{Election interference model}\label{sec:neq}
We consider an election between two candidates with no electoral 
institutions such as an Electoral College) 
We assume that the election process at any time $t \in [0, T]$ 
is represented by a public poll $Z_t \in [0,1]$.
The model is set in continuous time,
though when we estimate parameters statistically in Sec.\ \ref{sec:application} we move to a discrete-time 
analogue.
We hypothesize that the election dynamics take place in a latent space where dynamics are represented by 
$X_t \in \mathbb{R}$.
We will set $X_t < 0$ to be values of the latent poll that favor candidate A and $X_t > 0$ that favor 
candidate B.
The latent and observable space are related by 
$Z_t = \phi(X_t)$, where $\phi$ is a sigmoidal function which we choose to be $\phi(x) = \frac{1}{1 + e^{-x}}$.
(Any sigmoidal function that is bounded between zero and one will suffice and lead only to 
different parameter estimates in the context of statistical estimation.)
The actual result of the election---the number of votes that are earned by candidate B---is given by $\phi(X_T)$.
The election takes place in a population of $N$ voting agents.
Each voting agent updates their preferences over the candidates at each time step $t_n$ 
by a random variable $\xi_{n, t_n}$.
These random variables satisfy $E_n\left[ \xi_{n, t_k} \right] = 0$ and 
$E_n[\xi_{n,t_k}^2] < \infty$ for all $t$.
The increments of the election process are the sample means
of the voting agents' preferences at time $t$. 
In the absence of interference, the stochastic election model is an unbiased random walk:
\begin{equation}\label{eq:elec-process-discrete}
	X_{t_{k+1}} = X_{t_k} + \frac{1}{N}\sum_{1 \leq n \leq N} \xi_{n,t_k} \Delta t_k.
\end{equation}
where we have put $\Delta t_k = t_{k+1} - t_{k}$.
\begin{figure*}
\centering
	\includegraphics[width=\textwidth]{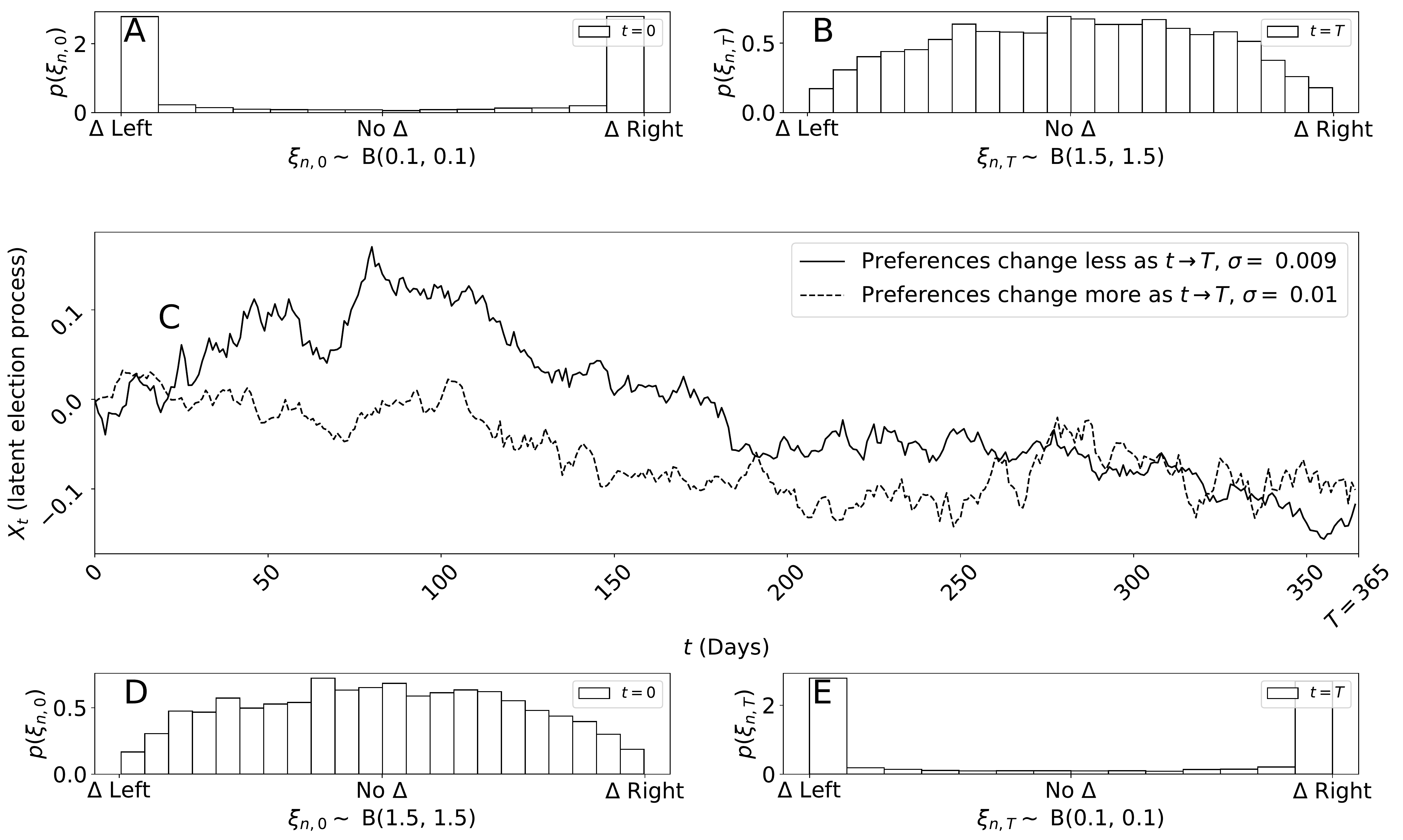}
	\caption{
	The random walk latent space election model is an accurate approximation to 
	multiple different population candidate preference updates. 
	The latent election process evolves according to 
	$X_{k+1} = X_{k} + \frac{1}{N}\sum_{1 \leq n \leq N} \xi_{n,k}$. 
	The random variable $\xi_{n,k}$ is voting agent $n$'s shift toward the left or right. 
	of the political spectrum at time $k$.
	In the center panel, 
	the solid curve is a draw from the latent election process resulting from 
	the preference updates $\xi_{n,t} \sim B\left( 0.1\frac{T - t}{T} + 1.5 \frac{t}{T},
	0.1\frac{T - t}{T} + 1.5 \frac{t}{T}\right)$, where $B(\alpha, \beta)$ is the Beta
	distribution and we have set $T = 365$.
	As $t \rightarrow T$, the electorate exhibits
	increasing resistance to change 
	in their political viewpoints.
	We display the preference shift distributions at $t = 0$ in Panel A and at $t = T$ in Panel B.
	For contrast, the dashed curve is a draw from the latent election 
	process resulting from $\xi_{n,t} \sim B\left( 1.5\frac{T - k}{T} + 0.1\frac{k}{T},
	1.5\frac{T - k}{T} + 0.1 \frac{k}{T}\right)$, 
	which describes an electorate in which agents more often have changing political preferences 
	as $t \rightarrow T$.
	We show the 
	corresponding preference shift distributions at $t = 0$ in Panel D and at $t = T$ in Panel E.
	Despite these preference updates that are, in some sense, opposites of each other, the latent processes
	$X_t$ are statistically very similar and are both well-modeled by the continuum approximation 
	$\dee X_t = \sigma \dee \wiener_t$.
	}
	\label{fig:process-creation}
\end{figure*}
We display sample realizations of this process for different distributions of 
$\xi_{n,t_k}$ in Fig.\ \ref{fig:process-creation}.
Though one distribution of preference changes has a larger variance than the other,
the sample paths of $X_{t_k}$ are statistically
similar for each since $\frac{1}{N}\sum_n \xi_{n,t_k}$ does not vary much between the distributions.
When $N$ is large we can reasonably approximate this discrete agent electoral process by a Wiener process, 
$\dee X_t = \sigma \dee \wiener_t$,
where $\sigma^2 \approx \text{Var}\left( \frac{1}{N}\sum_{1 \leq n \leq N} \xi_{n,t} \right)$, 
This limit is valid in the limit of large $N$.
\medskip

\noindent
If the preference change random variables $\xi_{n,k}$ did not satisfy $E_n[\xi_{n,k}] = 0$
then the random walk approximation to Eq.\ \ref{eq:elec-process-discrete} would not necessarily be valid.
For example, if $\{\xi_{n,k}\}_{k \geq 0}$ were a random walk or were trend-stationary
for each $n$,
then $\{E_n[\xi_{n,k}]\}_{k \geq 0}$ would also respectively be a random walk or trend-stationary. 
A trend-stationary univariate time series is a stochastic process $x_t = f(t) + \eps_t$,
where $\eps_t$ is a stationary process and $f(\cdot)$ is a deterministic function of time
\cite{nelson1982trends,dejong1992integration}.
A univariate time series with a unit root is a time series that can be written as an 
autoregressive process of order $p$ ($AR(p)$), 
$x_t - \sum_{p'=1}^p \beta_{p'} x_{t - p'} = \eps_t$, such that
the polynomial $z^p - \sum_{p'=1}^p \beta_{p'} z^{t - p'} = 0$ 
has a root on the unit circle when solved
over the complex numbers.
A random walk is a special case of the $AR(p)$ process with characteristic polynomial given by $z - 1 = 0$,
which has the unit root $z = 1$.
Trend-stationary and unit-root time series differ fundamentally in that a trend-stationary process subjected to 
an exogenous shock will eventually revert to its mean function $f(t)$.
This is not the case for a stochastic process with a unit root.
A unit root or trend-stationary $\xi_{n,k}$ would model a population in which political preferences were undergoing
a shift in population mean rather than just in individual preferences.
\medskip

\noindent
However, there do exist cases where the random walk approximation is valid even when $E_n[\xi_{n,k}] \neq 0$.
If the stochastic 
evolution equation for $E_n[\xi_{n,k}]$ has a stationary colored noise with exponentially-decaying 
covariance function as its solution,
then the integral of this noise satisfies a Stratonovich-type equation 
\cite{gardiner1985handbook,jung1987dynamical,hanggi1995colored}.
This equation would be a generalized 
version of the basic random walk model considered here, but
we will not consider this scenario in the remainder of this work.
\medskip

\noindent
We denote the control policies of Red and Blue (the functions by which Red and Blue attempt to influence 
(or prevent influence on) the election) by $u_{\rr}(t)$ and $u_{\bb}(t)$.
These functions are one-dimensional continuous-time stochastic processes (time series).
The term ``policy'' originates
from the fields of economics and reinforcement learning
\cite{bellman1954dynamic,bellman1966dynamic,sutton2018reinforcement}.
These control policies are abstract variables in the context of our model, but 
we interpret them as expenditures on interference operations.
We assume that Red and Blue can affect the mean trajectory of the election but not its volatility (standard 
deviation
of its increments).
We make this assumption because $X_t$ is an approximation to the process described by 
Eq.\ \ref{eq:elec-process-discrete}.
As we show in Fig.\ \ref{fig:process-creation},
the variance of the electoral process does not change much even when the voting population's underlying 
preference change distributions have differing variance and kurtosis.
Under the influence of Red's and Blue's control policies, the election dynamics become
\begin{equation}
	\dee X_t = F(u_{\rr}(t), u_{\bb}(t))\ \dee t + \sigma\ \dee \wiener_t,\ X_0 = y.
\end{equation}
The function $F$ captures the mechanism by which Red and Blue affect the mean dynamics of the latent electoral process.
We assume that $F$ is at least twice continuously-differentiable for connvenience.
To first order expansion we have $F(u_{\rr}(t), u_{\bb}(t)) = a_0 + a_{\rr} u_{\rr}(t) + a_{\bb} u_{\bb}(t) + \bigoh(u^2)$,
which is most accurate near $u = 0$.
We approximate the state equation by
\begin{equation}\label{eq:state}
	\dee X_t = [u_{\rr}(t) + u_{\bb}(t)]\ \dee t + \sigma\ \dee \wiener_t,\ X_0 = y,
\end{equation}
since we have assumed zero endogenous drift and can absorb constants into the definition of the control
policies.
We will use Eq.\ \ref{eq:state} as the state equation for the remainder of the paper.

\subsection{Subgame-perfect Nash equilibria}\label{sec:sgpe}
Red and Blue each seek to minimize separate scalar cost functionals
of their own control policy and the other agent's control
policy.
We will assume that the agents do not incur a running cost from the value of the state variable, 
although we will revisit this assumption in Sec.\ \ref{sec:discussion}.
The cost functionals are therefore 
\begin{equation}
	E_{u_{\rr}, u_{\bb}, X}\left\{ \Phi_{\rr}(X_T) + 
	\int_0^T C_{\rr}(u_{\rr}(t), u_{\bb}(t))\ \dee t \right\}\label{eq:red-cost},
\end{equation}
and
\begin{equation}
	E_{u_{\rr}, u_{\bb}, X}\left\{ \Phi_{\bb}(X_T) + \int_0^T C_{\bb}(u_{\rr}(t), u_{\bb}(t))\ \dee t \right\}\label{eq:blue-cost}.
\end{equation}
The functions $C_{\rr}$ and $C_{\bb}$ represent the running cost or benefit of conducting 
election interference operations.
We assume the cost functions have the form
\begin{equation}\label{eq:cost-function}
	C_i(u_{\rr}, u_{\bb}) = u_i^2 - \lambda_i u_{\lnot i}^2
\end{equation}
for $i \in \{\rr, \bb\}$.
The notation $\lnot i$ indicates the set of all other players.
For example,
if $i = \rr$, $\lnot i = \bb$.
This notation originates in the study 
of noncooperative economic games.
The non-negative scalar $\lambda_i$ parameterizes the utility gained by player $i$ from observing 
player $\lnot i$'s effort.
If $\lambda_i > 0$, player $i$ gains utility from player $\lnot i$'s expending
resources, while if $\lambda_i = 0$, player $i$ has no regard for $\lnot i$'s level of effort but only for
their own running cost and the final cost.
Our assumption that cost accumulates quadratically with magnitude of the control policy 
is common in optimal control theory 
\cite{kappen2007introduction,aastrom2012introduction,georgiou2013separation}.
We can justify the functional form of Eq.\ \ref{eq:cost-function} as follows.
Suppose that an arbitrary analytic cost function for player $i$ as 
$\mathcal{C}_i(u_\rr, u_\bb) = \mathcal{C}^{(i)}(u_i) - \lambda_i \mathcal{C}^{(\lnot i)}(u_{\lnot i})$.
We make the following assumptions:
\begin{itemize}
	\item It is equally costly to for player $i$ to conduct operations that favor candidate A or candidate B.
		This imposes the constraint that $\mathcal{C}^{(i)}$ and $\mathcal{C}^{(\lnot i)}$ are even functions.
	\item Player $i$ conducting no interference operations results in player $i$'s incurring no direct cost from this 
	choice. In other words, if $u_i(t) = 0$ at some $t$, player $i$ does not incur any cost from this.
\end{itemize}
With these assumptions, the 
first non-zero term in the Taylor expansion of $\mathcal{C}_i$ is given by Eq.\ \ref{eq:cost-function}.

\subsubsection{Choice of final conditions}
Finding optimal play in noncooperative games often requires solving the game backward through time
\cite{zermelo1913anwendung,nash1951non,fudenberg1989noncooperative,mas1995microeconomic}.
Therefore, we must define final conditions that specify the cost that Red and Blue incur from the 
actual election result $\phi(X_T)$.
Red and Blue might have different final conditions
because of their qualitatively distinct objectives.
Since Red wants to influence the outcome of the election in Blue's country in favor of candidate A, 
their final cost function $\Phi_{\rr}$ must satisfy $\Phi_{\rr}(x) < \Phi_{\rr}(y)$ for all $x < 0$ and $y >0$. 
In the final conditions that we are about to present, we also assume that $\Phi_{\rr}$
is monotonically non-decreasing everywhere.
We relax these assumption in Sec.\ \ref{sec:application} when we
confront this model with election interference-related data.
To the extent that this model describes reality,
it is probably not true that these restrictive assumptions
on the final condition are always satisfied.
However, one simple final condition that satisfies these requirements is $\Phi_{\rr}(x) = c_0 + c_1x$,
but this allows the unrealistic limiting condition of infinite benefit if 
candidate A gets 100\% of the vote in the election
and infinite cost if candidate A gets 0\% of the vote.
We will also consider two Red final conditions with cost that remains bounded as $x \rightarrow \pm \infty$: one smooth, 
$\Phi_{\rr}(x) = \tanh(x)$; and one discontinuous, $\Phi_{\rr}(x) = \Theta(x) - \Theta(-x)$.
By $\Theta(\cdot)$ we mean the Heaviside step function.
\medskip

\noindent
Blue wants to reduce the overall impact of Red's interference 
operations on the electoral process.
Since Blue is \textit{a priori}
indifferent between the outcomes of the election,
it initially seems that 
$\Phi_{\bb}(x) = 0$.
However, if $\lambda_{\bb} = 0$, this results in Blue taking no action due to the 
functional form of Eq.\ \ref{eq:cost-function}.
In other words, if Blue does not gain utility from Red expending resources, then Blue will not try to stop 
Red from interfering in an election in Blue's country.
Hence we believe that Blue cannot be indifferent about the election outcome.
\medskip

\noindent
We present three possible final conditions representing Blue's preferences over the election result. 
Blue might believe that a result was due to Red's interference if $X_T$ is too far from 
$E_0[X_T] = 0$. 
An example of a smooth function that represents this belief is 
$\Phi_{\bb}(x) = \frac{1}{2}x^2$.
However, this neglects the reality that Red's objective is not to have either candidate A or candidate B win by a
large margin, but rather to have candidate A win, i.e., have $X_T < 0$. Thus Blue might be unconcerned 
about larger positive values of the state variable and, modifying the previous function suitably, 
have $\Phi_{\bb}(x) = \frac{1}{2}x^2 \Theta(-x)$.
Alternatively, Blue may accept the result of the election if it does not deviate ``too far'' from the 
initial expected value. An example of a discontinuous final condition that represents these preferences is
$\Phi_{\bb}(x) = \Theta(|x| - \Delta) - \Theta(\Delta - |x|)$,
where $\Delta > 0$ is Blue's accepted margin of error.
\medskip

\noindent
Though we could propose many other possible final conditions, these example functions demonstrate
some possible payoff structures. 
We include:
\begin{itemize}
	\item ``First-order'' functions that could result from the Taylor expansion about zero of an arbitrary 
analytic final condition.
These functions are linear, in the case of Red's antisymmetric final condition, and quadratic in the case 
of Blue's smooth symmetric final condition (which is the first non-constant term in the Taylor expansion of an 
even analytic function); 
\item
Smooth functions that represent bounded 
preferences over the result of the electoral process and the recognition that Red
favors one candidate in particular; and 
\item
Discontinuous final conditions that model ``all-or-nothing'' preferences over the outcome (either candidate A 
wins or they do not; either Red interferes less than a certain amount or they interfere more).
\end{itemize}
These functions do not capture some behavior that might exist in real election interference operations. 
For example, Red's preferences could be as follows:
``we would prefer that candidate A wins the election, 
but if they cannot, then we would like candidate B to win by a landslide so that we can 
claim the electoral system in Blue's country was rigged against candidate A''.
These preferences correspond to a final condition with a global minimum at some $x < 0$ but a secondary local 
minimum at $x \gg 0$.
This situation is not modeled by any of the final conditions that we have stated.
In Sec.\ \ref{sec:application} we relax the assumption that the final conditions are parameterized according to 
any of the functional forms considered in thise section and instead 
infer them from observed election and election interference proxy data
using the method described in Sec.\ \ref{sec:inference}.

\subsubsection{Value functions}

Applying the dynamic programming principle \cite{bellman1954dynamic,bellman1966dynamic} to 
Eqs.\ \ref{eq:state}, \ref{eq:red-cost}, and \ref{eq:blue-cost} leads to a system of coupled 
Hamilton-Jacobi-Bellman equations for the Red and Blue value functions,
\begin{equation}
	-\frac{\partial V_{\rr}}{\partial t} =
	\min_{u_{\rr}} \left
	\{ \frac{\partial V_{\rr}}{\partial x}[u_{\rr} + u_{\bb}] + u_{\rr}^2 - \lambda_{\rr} u_{\bb}^2 
	+ \frac{\sigma^2}{2}\frac{\partial^2 V_{\rr}}{\partial x^2} \right\} \label{eq:valred},
\end{equation}
and
\begin{equation}
	-\frac{\partial V_{\bb}}{\partial t} =
	\min_{u_{\bb}}\left\{ \frac{\partial V_{\bb}}{\partial x}[u_{\rr} + u_{\bb}] + u_{\bb}^2 - \lambda_{\bb} u_{\rr}^2 
	+\frac{\sigma^2}{2}\frac{\partial^2V_{\bb}}{\partial x^2}\right\}\label{eq:valblue}.
\end{equation}
The dynamic programming principle does not result in an Isaacs equation because the game is not zero-sum and 
the cost functionals for Red and Blue can have different functional forms.
(The Isaacs equation is a nonlinear elliptic or parabolic equation that arises in the study of two-player,
zero-sum games in which one player 
attempts to maximize a functional and the other player attempts to minimize it 
\cite{buckdahn2008stochastic,pham2014two}.)
Performing the minimization with respect to the control variables gives the Nash equilibrium control policies,
\begin{align}
	u_{\rr}(t) &= -\frac{1}{2}\frac{\partial V_{\rr}}{\partial x}\bigg|_{(t, X_t)}\label{eq:red-control}\\
	u_{\bb}(t) &= - \frac{1}{2}\frac{\partial V_{\bb}}{\partial x}\bigg|_{(t, X_t)}\label{eq:blue-control},
\end{align}
and the exact functional forms of Eqs.\ \ref{eq:valred} and \ref{eq:valblue},
\begin{widetext}
\begin{align}
	-\frac{\partial V_{\rr}}{\partial t} &= 
	-\frac{1}{4}\left( \frac{\partial V_{\rr}}{\partial x} \right)^2 -
	\frac{1}{2}\frac{\partial V_{\rr}}{\partial x}\frac{\partial V_{\bb}}{\partial x}
	- \frac{\lambda_{\rr}}{4}\left(\frac{\partial V_{\bb}}{\partial x}\right)^2
	+ \frac{\sigma^2}{2}\frac{\partial^2 V_{\rr}}{\partial x^2},\qquad V_{\rr}(x,T) = \Phi_{\rr}(x);
	\label{eq:red-value}\\
	-\frac{\partial V_{\bb}}{\partial t} &= 
	-\frac{1}{4}\left( \frac{\partial V_{\bb}}{\partial x} \right)^2 -
	\frac{1}{2}\frac{\partial V_{\bb}}{\partial x}\frac{\partial V_{\rr}}{\partial x}
	- \frac{\lambda_{\bb}}{4}\left(\frac{\partial V_{\rr}}{\partial x}\right)^2
	+ \frac{\sigma^2}{2}\frac{\partial^2 V_{\bb}}{\partial x^2},\qquad V_{\bb}(x,T) = \Phi_{\bb}(x).
	\label{eq:blue-value}
\end{align}
\end{widetext}
When solved over the entirety of state space,
solutions to Eqs.\ \ref{eq:red-value} and \ref{eq:blue-value} constitute the strategies of a subgame-perfect 
Nash equilibrium.
No matter the action taken by player $\lnot i$ at time $t$, player $i$ is able to 
respond with the optimal action at time $t + \dee t$.
This is the (admittedly-informal) definition 
of a subgame-perfect Nash equilibrium in a continuous-time differential game \cite{mas1995microeconomic}.
Given the solution pair $V_{\rr}(x,t)$ and $V_{\bb}(x,t)$, we can
write the distribution of $x$, $u_{\rr}$, and  $u_{\bb}$ analytically.
Substitution of Eqs.\ \ref{eq:red-control} and \ref{eq:blue-control} into Eq.\ \ref{eq:state} gives 
$\dee x = -\frac{1}{2}\left\{ \frac{\partial V_{\rr}}{\partial x}|_{(t, x)}
+ \frac{\partial V_{\bb}}{\partial x}|_{(t, x)}\
\right\}\dee t + \sigma \dee \wiener$. 
We discretize this equation over $N$ timepoints to obtain 
\begin{equation}
	\begin{aligned}
		&x_{n + 1} - x_n + \frac{\Delta t}{2}\left[ V_{\rr n}' + V_{\bb n}' \right]\\
	&\qquad
	- (\Delta t)^{1/2}\sigma w_n - y\delta_{n,0} = 0,
	\end{aligned}
\end{equation}
with $w_n \sim \mathcal{N}(0,1)$, $\Delta t = t_{n+1} - t_n$, and $n = 0,...,N - 1$
and where we have put $V'_{in} \equiv V_i'(x_n, t_n)$.
Thus the distribution of an increment of the latent electoral process is 
\begin{equation}
	p(x_{n + 1} | x_{n}) = \frac{1}{\sqrt{2\pi\sigma^2 \Delta t}}e^{-\frac{\Delta t}{2\sigma^2}
	(\frac{x_{n + 1} - x_n}{\Delta t} + \frac{1}{2}[V'_{\rr n} + V'_{\bb n}] - y\frac{\delta_{n0}}{\Delta t})^2}.
\end{equation}
Now, using the Markov property of $X_t$, we have
\begin{align}
	p(x_1,...,x_{N}|x_0) &= \prod_{n=0}^{N - 1} p(x_{n + 1}|x_n)\\
		 &= 
		\frac{1}{(2\pi\sigma^2\Delta t)^{N/2}}\exp\left\{ -\frac{1}{2\sigma^2} S(x_1,...,x_N) \right\} 
		\label{eq:state-distribution},
\end{align}
where
\begin{equation}
	\begin{aligned}
	S(x_1,...,x_N) &= \sum_{n=0}^{N-1}\Delta t\Big[ \frac{x_{n+1} - x_n}{\Delta t} \\ 
		&\quad+ \frac{1}{2}[V_{\rr n}' + V_{\bb n}'] - \frac{y \delta_{n,0}}{\Delta t}\Big]^2.
	\end{aligned}
\end{equation}
Taking $N \rightarrow \infty$ as $N \Delta t = T$ remains constant gives a functional Gaussian distribution,
\begin{equation}\label{eq:gaussian-path-distribution}
	p(x(0\rightarrow T)|x_0) = \frac{1}{\mathcal{Z}}\exp \left\{ -\frac{1}{2\sigma^2} S[x(0\rightarrow T)] \right\},
\end{equation}
with action 
\begin{equation}
	\begin{aligned}
		S[x(0\rightarrow T)] &= \int_0^T \Big[ \frac{\dee x}{\dee t} 
	+ \frac{1}{2}\Big( \frac{\partial V_\rr}{\partial x}\Big|_{x = x(t)}
		\frac{\partial V_\bb}{\partial x}\Big|_{x = x(t)}
	\Big) \\
		&\quad- y\delta(t - t_0) \Big]^2\ \dee t
	\end{aligned}
\end{equation}
and partition function 
\begin{equation}
	\mathcal{Z} = \int_{x(0)}^{x(T)}\mathcal{D}x(0 \rightarrow T)
	\exp\left\{ -\frac{1}{2\sigma^2}S[x(0\rightarrow T)]\right\}. 
\end{equation}
We have denoted by $x(s \rightarrow t)$ the actual 
path followed by the latent state from time $s$ to time $t$.
The measure $\mathcal{D}x(0\rightarrow T)$ is classical Wiener measure.
Since $u_{\rr}(0\rightarrow T)$ and $u_{\bb}(0\rightarrow T)$ are
deterministic time-dependent functions of $x(0 \rightarrow T)$, we can find their distributions 
explicitly using the probability distribution Eq.\ \ref{eq:state-distribution} and 
the appropriate time-dependent Jacobian transformation. 
These analytical results are of limited utility because we 
are unaware of analytical solutions to the system given in Eqs. \ref{eq:red-value} and \ref{eq:blue-value},
and hence $V_{\rr}'(x,t)$ and $V_{\bb}'(x,t)$ must be approximated.
In Sec.\ \ref{sec:credible} we will derive analytical results that are valid when 
player $i$ announces a credible commitment to a particular control path.
\medskip

\begin{figure*}
\centering
	\includegraphics[width=\textwidth]{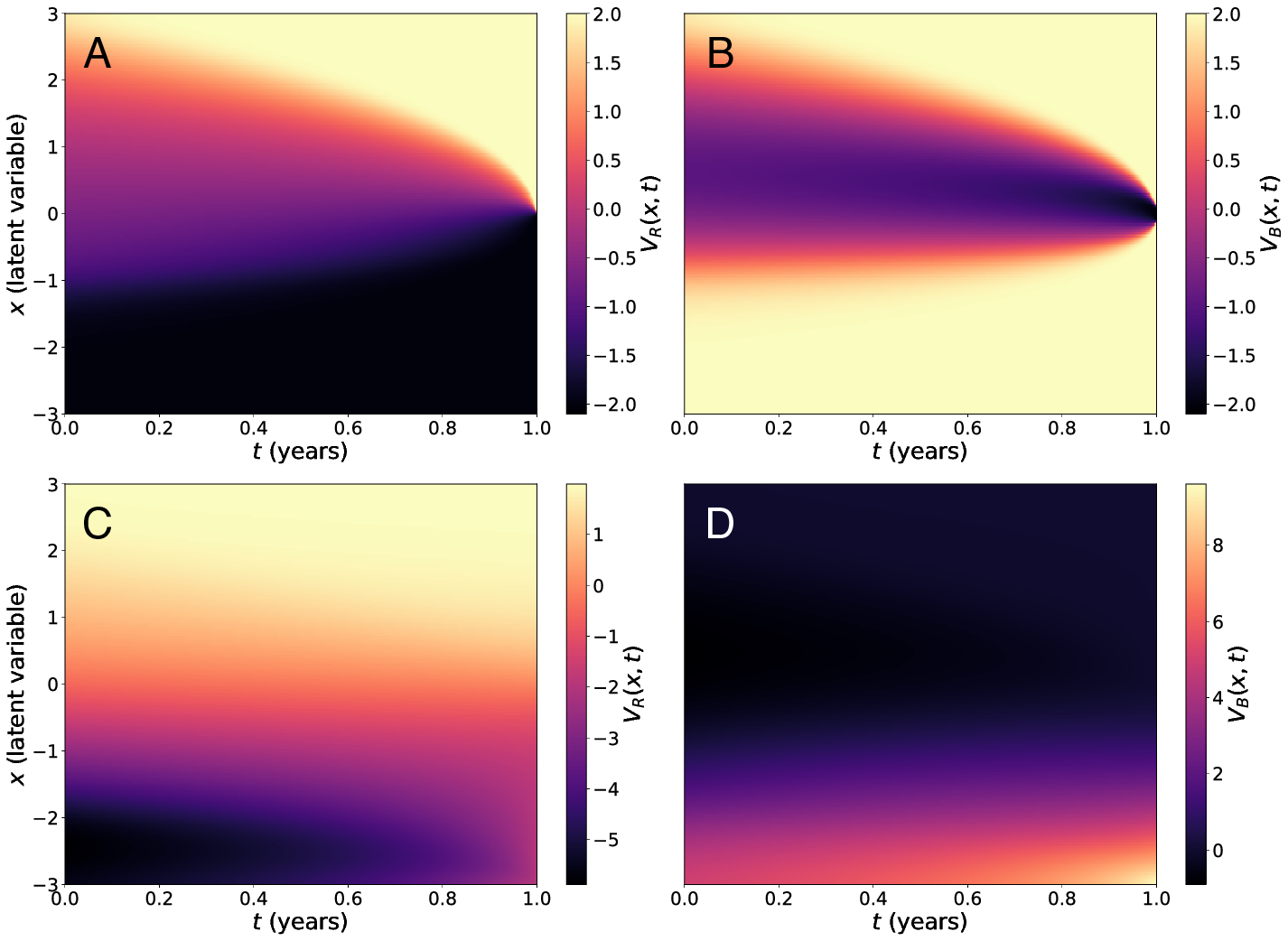}
	\caption{
		Example value functions corresponding to the system Eqs.\ \ref{eq:red-value} and 
	\ref{eq:blue-value}.
	Panels A and B display $V_{\rr}(x,t)$ and $V_{\bb}(x,t)$ respectively for 
	$\lambda_{\rr} = \lambda_{\bb} = 0$, $\Phi_{\rr}(x) = 2[\Theta(x) - \Theta(-x)]$,
	and $\Phi_{\bb}(x) = 2[\Theta(|x| - 0.1) - \Theta( 0.1 - |x|)]$ with $\Delta = 0.1$, while
	panels C and D display $V_{\rr}(x,t)$ and $V_{\bb}(x,t)$ respectively for 
	$\lambda_{\rr} = \lambda_{\bb} = 2$, $\Phi_{\rr}(x) = 2 \tanh(x)$, and 
	$\Phi_{\bb}(x) = \frac{1}{2}x^2\Theta(-x)$.
	For each solution we enforce Neumann no-flux boundary conditions and set $\sigma = 0.6$. 
	The solution is computed on a grid with $x \in [-3, 3]$, setting $\dee x = 0.025$, and 
	integrating for $N_t = $ 8,000 timesteps.
	}
	\label{fig:example-values}
\end{figure*}

\noindent
We find the value functions $V_{\rr}(x,t)$ and $V_{\bb}(x,t)$ numerically
through
backward iteration, enforcing a Neumann boundary condition at $x = \pm 3$, which corresponds to 
bounding polling popularity of candidate B from below by $100 \times \phi(-3) = 4.7\%$ and from above by 
$100 \times \phi(3) = 95.3\%$
\footnote{
	Code to recreate simulations and plots in this 
	paper, or to create new simulations and ``what-if'' scenarios, is 
	located at 
	\href{https://gitlab.com/daviddewhurst/red-blue-game}{https://gitlab.com/daviddewhurst/red-blue-game}.
}.
We display example realizations of the value functions for different 
$\lambda_i$ and final conditions in Fig.\ \ref{fig:example-values}.
The value functions display diffusive dynamics because the state equation is driven by Gaussian white noise.
The value functions also depend crucially on the final condition.
When the final conditions are discontinuous (as in the top panels of Fig.\ \ref{fig:example-values})
the derivatives of the value function reach larger magnitudes and vary more rapidly than 
when the final conditions are continuous.
This has 
consequences for the game-theoretic interpretation of these results, 
as we discuss 
in Sec.\ \ref{sec:parameters}.
Fig.\ \ref{fig:example-values} also demonstrates that the 
extrema of the value functions are not as large in magnitude when $\lambda_{\rr} = \lambda_{\bb} = 0$ as 
when $\lambda_{\rr} = \lambda_{\bb} = 2$; this is 
because higher values of $\lambda_i$ mean that player $i$ derives 
utility not only from the final outcome of the game but also from causing player $\lnot i$ to expend resources
in the game.
\medskip

\begin{figure}
\centering
	\includegraphics[width=\columnwidth]{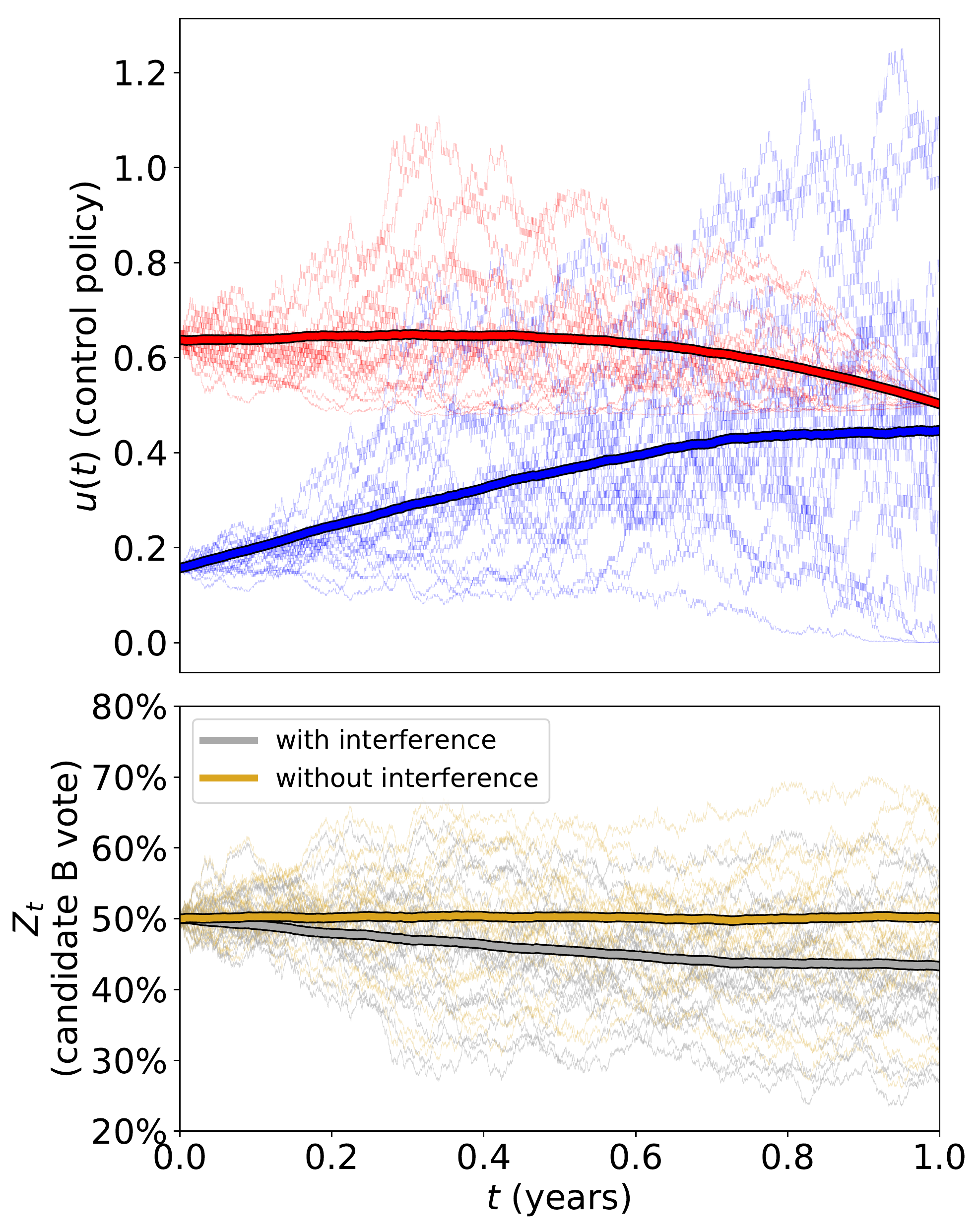}
	\caption{We display realizations of $u_{\rr}$ and $u_{\bb}$ in the top panel 
	and paths of the electoral process in the bottom panel. 
	We draw these realizations from the game simulated with parameters 
	$\lambda_{\rr} = \lambda_{\bb} = 2$, $\Phi_{\rr}(x) = x$, and $\Phi_{\bb}(x) = \frac{1}{2}x^2\Theta(-x)$. 
	For this parameter set, Blue is fighting a losing battle since optimal play by both players 
	results in lower $E[Z_t]$ than for the electoral process without any interference.
	}
	\label{fig:red-blue-control-example}
\end{figure}

\noindent
Eqs.\ \ref{eq:valred} and \ref{eq:valblue} give the closed-loop control 
policies given the current state $X_t$ and 
time $t$, $u_{\rr}$ and $u_{\bb}$.
We display examples of $u_{\rr}$, $u_{\bb}$ and the electoral process $Z_t$ in Fig.\  
\ref{fig:red-blue-control-example}.
For this example, we simulate the game with parameters 
$\lambda_{\rr} = \lambda_{\bb} = 2$, $\Phi_{\rr}(x) = x$, and 
$\Phi_{\bb}(x) = \frac{1}{2}x^2\Theta(-x)$.
We plot the control policies in the top panel.
The mean control policies 
$E[u_{\rr}]$ and $E[u_{\bb}]$ are displayed in thicker curves.
For this parameter set,
it is optimal for Red to begin play with a larger amount of interference than Blue does and on average
decrease the level of interference over time.
Throughout the game Blue increases their resistance to Red's interference.
Even though Blue resists Red's interference, Red is able to accomplish 
their objective of causing candidate A to win.

\subsubsection{Inference and prediction}\label{sec:inference}

The solutions to Eqs.\ \ref{eq:red-value} and \ref{eq:blue-value} are functions of the final conditions 
$V_\rr(x,T) = \Phi_\rr(x)$ and $V_\bb(x,T) = \Phi_\bb(x)$.
It is possible to perform both inference and prediction at times $t < T$ even 
when $\Phi_\rr(x)$ and $\Phi_\bb(x)$ are not known.
To do this, we assume that the system given by Eqs.\ \ref{eq:red-value} and \ref{eq:blue-value} 
has a unique solution given particular final conditions $\Phi_\rr$ and $\Phi_\bb$.
Though we have numerical evidence to suggest that such solutions do exist and are unique,
we have not proved that this is the case.
In inference, we want to find the distributions of values of some unobserved parameters of the system.
We will suppose that we want to infer $\Phi_\rr$ and $\Phi_\bb$ given the observed paths
$x(0\rightarrow t), u_\rr(0\rightarrow t), u_\bb(0\rightarrow t)$ with $t < T$.
For simplicity we assume that we know all other parameters of Eqs.\ \ref{eq:red-value} and \ref{eq:blue-value}
with certainty.
Then the posterior distribution of $\Phi_\rr$ and $\Phi_\bb$ reads
\begin{equation}\label{eq:final-cond-inference}
	p(\Phi_\rr, \Phi_\bb|X_s) \propto p(x(0 \rightarrow t) | \Phi_\rr, \Phi_\bb)\ p(\Phi_\rr, \Phi_\bb).
\end{equation}
The likelihood $p(x(0 \rightarrow t) | \Phi_\rr, \Phi_\bb)$ 
is Gaussian, as shown in Eq.\ \ref{eq:gaussian-path-distribution}, 
and depends on the time-dependent Jacobian transformtions
defined implicitly by the solutions of Eqs.\ \ref{eq:red-value} and \ref{eq:blue-value}. 
The prior over final conditions $p(\Phi_\rr, \Phi_\bb)$ can be set proportional to unity if we want to use a 
maximum-likelihood approach and not account for our prior beliefs about the form of $\Phi_\rr$ and $\Phi_\bb$.
We can approximate $\Phi_\rr$ and $\Phi_\bb$ with functions parameterized by a 
finite set of parameters $a_{i,k}$,
where $i \in \{\rr, \bb\}$ and 
$k = 0,...,K$.
The functional prior $p(\Phi_\rr, \Phi_\bb)$ is then approximated by the multivariate distribution 
$p(k_{\rr,0},...,k_{\rr, K}, k_{\bb, 0},...,k_{\bb, K})$.
We will take this approach when performing inference in Sec.\ \ref{sec:application}.
\medskip

\noindent
We can predict future values of $x(t)$, and hence $u_\rr(t)$ and $u_\bb(t)$, similarly. 
Now we want to find the probability of observing $x(t \rightarrow T)$ given observed $x(0 \rightarrow t)$.
To do this, we integrate out all possible choices of $\Phi_\rr$ and $\Phi_\bb$ weighted by their posterior likelihood
given the observed path $x(0\rightarrow t)$.
The integration is taken with respect to a functional measure, $\mathcal{D}(\Phi_\rr(x), \Phi_\bb(x))$.
This means that the integration is taken over
all possible choices of $\Phi_\rr$ and $\Phi_\bb$ that lie in some particular class of functions
\footnote{
There is a large literature on nonparametric functional approximation that we cannot review here. 
Here is an example of this type of approximation.
If $\{x_k\}_{1\leq k \leq K}$ is any partition of $[a, b)$ and the vector
$(\Phi_\rr(x_k), \Phi_\bb(x_k))$ were jointly distributed Gaussian on $\mathbb{R}^{2K}$,
then we say that $\Phi_\rr$ and $\Phi_\bb$ are distributed according to a Gaussian process \cite{williams2006gaussian}.
Though this infinite-dimensional formalism can be useful when deriving theoretical results,
in practice we would approximate this process by a finite-dimensional multivariate Gaussian.
In Sec.\ \ref{sec:application} we 
actually approximate $\Phi_\rr$ and $\Phi_\bb$ by finite sums of Legendre polynomials and infer the coefficients
of these polynomials as our multivariate approximation.
}.
As in the case of inference, 
we can approximate $\Phi_\rr$ and $\Phi_\bb$ by functions parameterized by a finite set of 
parameters $a_{i,k}$ and integrate over the $2K$-dimensional domain of these parameters.
In the present work we do not predict any future values of the latent electoral process or control policies.

\subsubsection{Dependence of value functions on parameters}\label{sec:parameters}

\begin{figure*}
\centering
	\includegraphics[width=\textwidth]{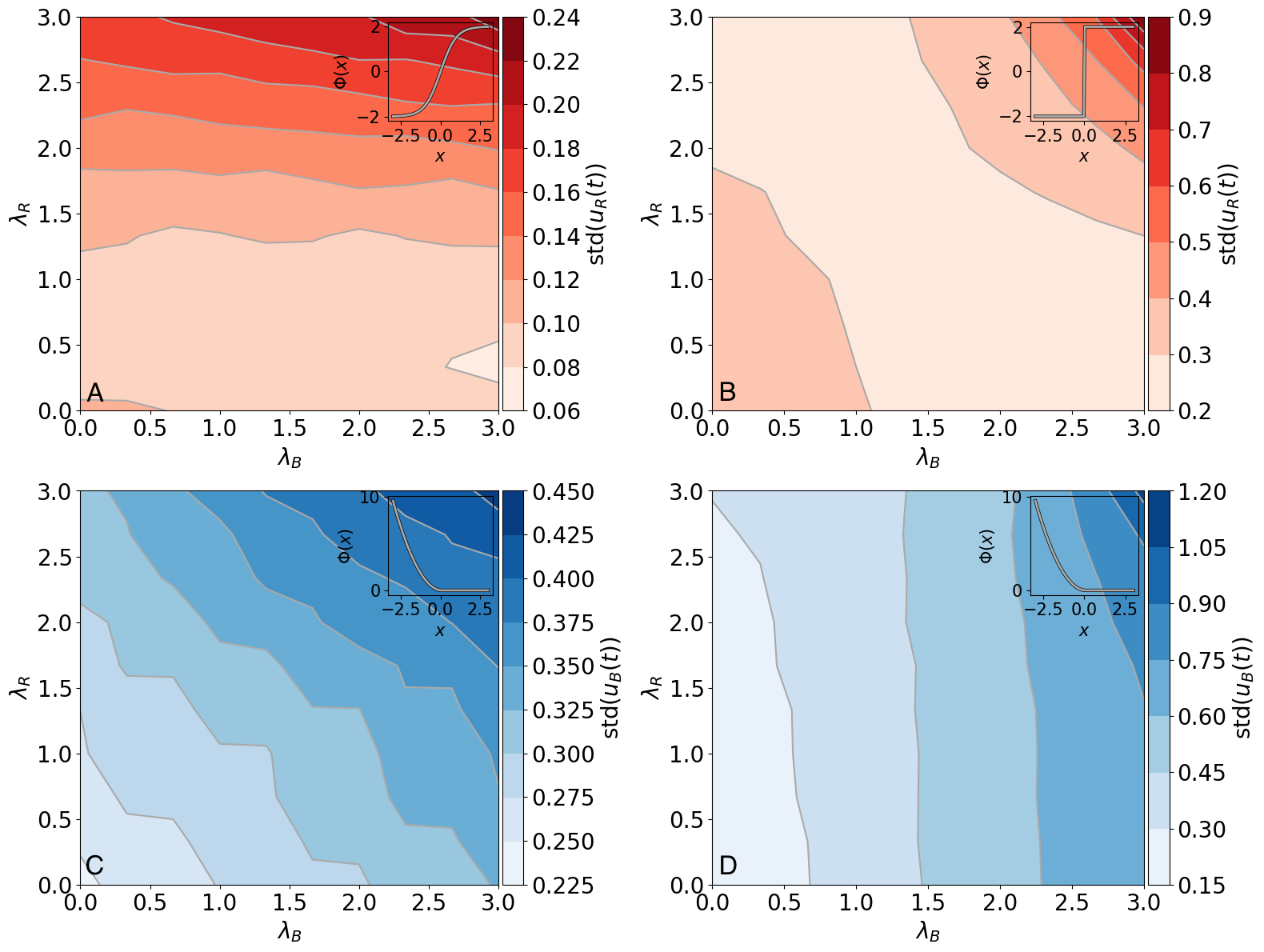}
	\caption{
		 Example sweeps over the coupling parameters 
		 $\lambda_{\rr}$ and $\lambda_{\bb}$ when Blue's final condition is set to 
		 $
		\Phi_{\bb}(x) = \frac{1}{2}x^2 \Theta(-x)$.
		 We vary the coupling parameters over $[0,3]$ and display the resulting 
		 standard deviation of the control
		 policies $u_{\rr}(x)$ and $u_{\bb}(x)$.
		 Panels A and B represent one coupled system of equations, while panels C and D represent a coupled system
		 of equations with a different set of final conditions.
		 In panel A, Red's value function is set to 
		 $\Phi_{\rr}(x) = \tanh(x)$, while in panel B it is given by 
		 $
		 \Phi_{\rr}(x) = \Theta(x) - \Theta(-x)$, where $\Theta(\cdot)$ is the Heaviside function.
		 We display a glyph of the corresponding final condition in the upper right corner of each panel.
		 Changing Red's continuous final condition $\tanh(x)$ to the discontinuous 
		 $\Theta(x) - \Theta(-x)$ results in substantially increased variation in the control policies of 
		 both players.
		 }
	\label{fig:lambda-sweep-montage}
\end{figure*}

\begin{figure*}
\centering
	\includegraphics[width=\textwidth]{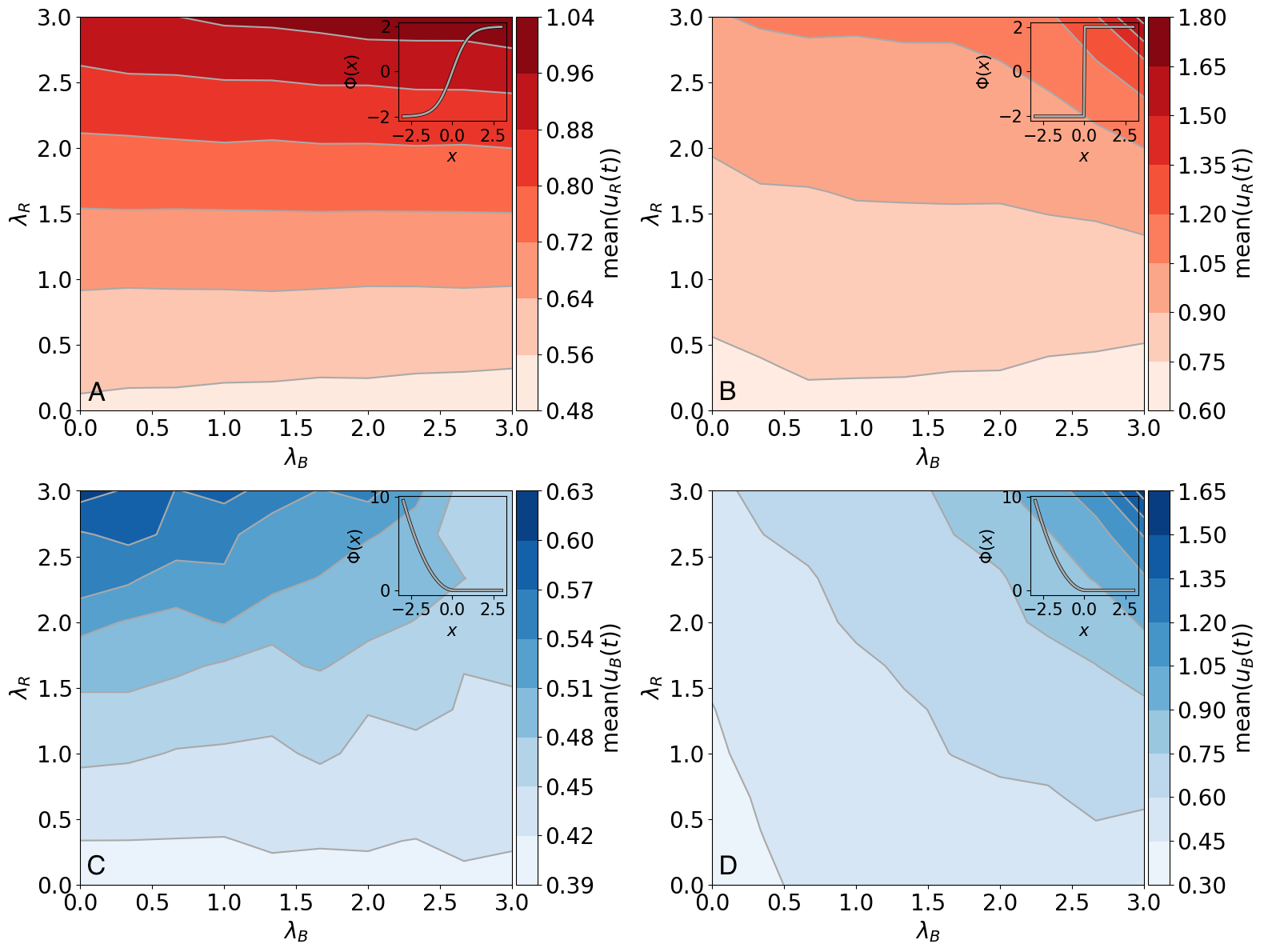}
	\caption{
		 Example sweeps over the coupling parameters 
		 $\lambda_{\rr}$ and $\lambda_{\bb}$ when Blue's final condition is set to 
		 $
		\Phi_{\bb}(x) = \frac{1}{2}x^2 \Theta(-x)$.
		 We vary the coupling parameters over $[0,3]$ and display the resulting 
		 means of the control
		 policies $u_{\rr}(x)$ and $u_{\bb}(x)$.
		 Panels A and B represent one coupled system of equations, while panels C and D represent a coupled system
		 of equations with a different set of final conditions.
		 In panel A, Red's value function is set to 
		 $\Phi_{\rr}(x) = \tanh(x)$, while in panel B it is given by 
		 $
		 \Phi_{\rr}(x) = \Theta(x) - \Theta(-x)$.
		 Altering Red's final condition from continuous to discontinuous causes a greater than 100\% increase 
		 in the maximum value of the mean of Blue's control policy.
		 }
		 \label{fig:lambda-sweep-montage-2}
\end{figure*}

We conducted a coarse parameter sweep over $\lambda_{\rr}$, $\lambda_{\bb}$, $\Phi_{\rr}$, and $\Phi_{\bb}$
to explore qualitative behavior of this game. 
We display the results of this parameter sweep for two combinations of final conditions
in Figs.\ \ref{fig:lambda-sweep-montage} and \ref{fig:lambda-sweep-montage-2}.
The upper right-hand corner of each panel of the figures displays the final condition of each player.
Holding Blue's final condition of $\Phi_{\bb}(x) = \frac{1}{2}x^2 \Theta(-x)$ constant, 
we compare the means and standard deviations of the Nash equilibrium strategies $u_{\rr}(t)$ and 
$u_{\bb}(t)$ across values of the coupling parameters $\lambda_{\rr}, \lambda_{\bb} \in [0,3]$
as Red's final condition changes from $\Phi_{\rr}(x) = \tanh(x)$ to 
$\Phi_{\rr}(x) = \Theta(x) - \Theta(-x)$.
\medskip

\begin{figure*}
\centering
	\includegraphics[width=\textwidth]{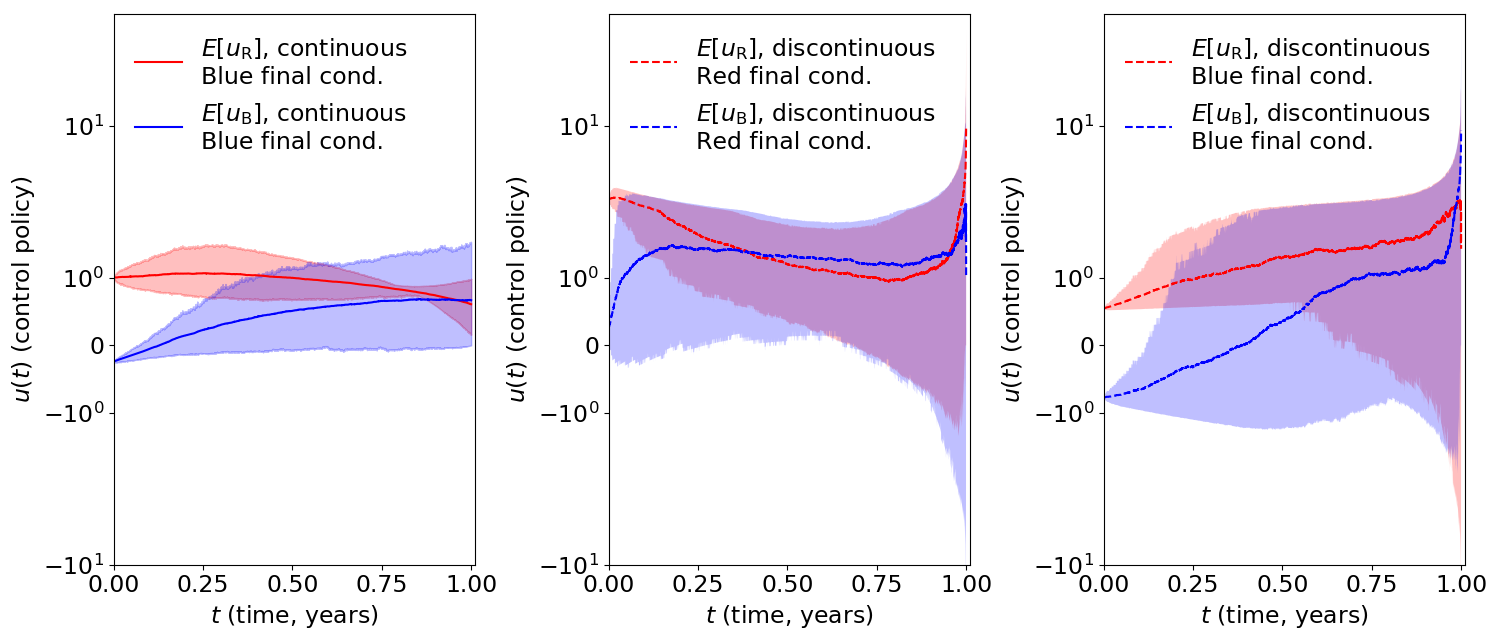}
	\caption{
		In the case of strong coupling ($\lambda_\rr$ and $\lambda_\bb \gg 0$), discontinuous final solutions by
		either player cause superexponential growth in the magnitude of each player's control policy.
		Here we set $\lambda_\rr = \lambda_\bb = 3$ and integrate three systems, varying only one final condition 
		in each.
		Panel A displays a system with two continuous final conditions: $\Phi_\rr(x) = \tanh(x)$ and 
		$\Phi_\bb(x) = \frac{1}{2}x^2\Theta(-x)$.
		Panel B displays the mean Red and Blue control policies when the Red final condition is changed to 
		$\Phi_\rr(x) = \Theta(x) - \Theta(-x)$ 
		as the Blue final condition remains equal to $\frac{1}{2}x^2\Theta(-x)$,
		while panel C shows the control policies when $\Phi_\bb(x) = \Theta(|x| > 1) - \Theta(|x| < 1)$
		and $\Phi_\rr(x) = \tanh(x)$.
		The shaded regions correspond to the middle 80 percentiles (10th to 90th percentiles) of $u_\rr(t)$ and 
		$u_\bb(t)$ for each $t$.
		When either player has a discontinuous final condition, the inter-percentile range is substantially wider
		for both players than when both players have continuous final conditions.
	}
	\label{fig:lambda-ts}
\end{figure*}

\noindent
For these combinations of final conditions, 
higher values of the coupling parameters $\lambda_i$ cause the  control policies to have higher variance. 
This increase in variance is more pronounced when Red's final condition is discontinuous, which is sensible 
since in this case $\lim_{t \rightarrow T^-}u_\rr(x, t) = -\frac{1}{2}\delta(x)$.
Appendix \ref{app:lambda-param-sweeps} contains similar figures for each $3^2 = 9$ combinations of Red example 
final conditions, $\Phi_\rr(x) \in \left\{ \tanh(x), \Theta(x) - \Theta(-x), x \right\}$ and 
Blue example final conditions, $\Phi_\bb(x) \in \left\{ \frac{1}{x}x^2, \frac{1}{2}x^2\Theta(-x),
\Theta(|x| - \Delta) - \Theta(\Delta - |x|) \right\}$.
We also find that certain combinations of parameters lead to an ``arms-race'' effect in both players' control policies.
For these parameter combinations, Nash equilibrium strategies entail
superexponential growth in the magnitude of each player's control policy near the end of the game.
Figure \ref{fig:lambda-ts} displays $E[u_\rr]$ and $E[u_\bb]$ for some of these parameter combinations, along with 
the middle 80\% credible interval (10th to 90th percentile) of $u_\rr(t)$ and $u_\bb(t)$ for each $t$.
A credible interval for the random variable $Y \sim p(y)$ is an interval into which $Y$ falls with a
particular probability \cite{chen1999monte}.
For example, the middle 80\% credible interval for $Y \sim p(y)$ is the interval $(a, b)$ for which 
$\int_a^b p(y)\ \dee y = 0.8$ and $\int_{-\infty}^a p(y)\ \dee y = \int_b^{\infty} p(y)\ \dee y = 0.1$.
\medskip

\noindent
This growth in the magnitude of each player's control policy occurs when either player has a discontinuous final 
condition. 
Although a discontinuous final condition by player $i$ leads to a greater increase in
mean magnitude of player $i$'s control policy than in player $\lnot i$'s,
the standard deviation of each player's policy 
exhibits similar superexponential growth.
To the extent that this model reflects reality, this points to a general statement about election interference
operations: 
an all-or-nothing mindset by either Red or Blue about the final outcome of the election 
leads to an arms race that negatively affects both players.
This is a general feature of any strategic interaction to which the model described by
Eqs.\ \ref{eq:state} - \ref{eq:blue-cost} applies.

\subsection{Credible commitment}\label{sec:credible}

If player $\lnot i$ credibly commits to playing a particular strategy $v(t)$ on all of $[0, T]$, 
then the problem of player $i$'s finding a subgame-perfect Nash equilibrium strategy profile becomes
an easier problem of optimal control. 
A credible commitment by player $\lnot i$ to the strategy $v(t)$ means that
\begin{itemize}
	\item player $\lnot i$ tells player $i$, either directly or indirectly, that 
		player $\lnot i$ will follow $v(t)$; and
	\item player $i$ should rationally believe that player $\lnot i$ will actually follow $v(t)$.
\end{itemize}
An example of a mechanism that makes commitment to a strategy credible
is the Soviet Union's ``Dead Hand'' automated second-strike nuclear response 
system.
This mechanism would launch a full-scale nuclear attack on the United States if it detected a nuclear strike on the 
Soviet Union \cite{yarynich2003c3,barrett2016false}.
The existence of this mechanism made the commitment to the strategy ``launch a full-scale nuclear attack on the United States given that any nuclear attack on my country has occurred'' credible even though the potential cost to both parties of 
executing the strategy was high.
\medskip

\noindent
When player $\lnot i$ credibly commits to playing $v(t)$,
player $i$'s problem reduces to finding the policy
$u(t)$ that minimizes the functional
\begin{equation}
	E_{u, X}\left\{ \Phi(X_T) + \int_0^T (u(t)^2 + \lambda v(t)^2)\ \dee t \right\},
\end{equation}
subject to the modified state equation 
\begin{equation}
	\dee x = [u(t) + v(t)]\dee t + \sigma \dee \wiener.
\end{equation}
Player $i$'s value function is now given by the solution to the HJB equation
\begin{equation}
-\frac{\partial V}{\partial t} =
	\min_{u} \left
	\{ \frac{\partial V}{\partial x}[u + v] + u^2 - \lambda v^2 
	+ \frac{\sigma^2}{2}\frac{\partial^2 V}{\partial x^2} \right\}.
\end{equation}
Performing the minimization gives the control policy $u(t) = - \frac{1}{2}\frac{\partial V}{\partial x}\Big|_{x=x(t)}$
and the explicit functional form of the HJB equation,
\begin{equation}\label{eq:commit-oc}
	\begin{aligned}
		-\frac{\partial V}{\partial t} &= -\frac{1}{4}\left( \frac{\partial V}{\partial x} \right)^2
	+ v(t) \frac{\partial V}{\partial x} + \lambda v(t)^2 
		+ \frac{\sigma^2}{2}\frac{\partial^2 V}{\partial x^2},\\
		V(x,T) &= \Phi(x).
	\end{aligned}
\end{equation}

\subsubsection{Path integral control}
\begin{figure}
\centering
	\includegraphics[width=\columnwidth]{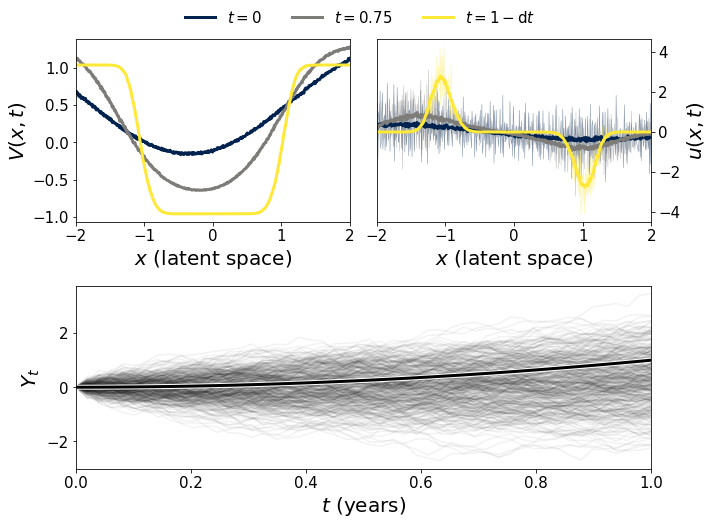}
	\caption{
	Result of the path integral Monte Carlo solution method applied to Eq.\ \ref{eq:commit-oc}
		with the final condition $\Phi(x) = \Theta(|x| > 1) - \Theta(|x| \leq 1)$ 
		and $v(t) = t^2$. 
	Approximate value functions are computed using $N = 10 000$ trajectories sampled from
	Eq.\ \ref{eq:abm-const-control} for each point $(x, t)$.
	We display approximate value functions in Panel A
	for $t \in \{0, 0.75, 1 - \dee t\}$ and 
	the corresponding approximate control policies $u(x,t)$ in Panel B, along with smoothed 
	versions of $u(x,t)$, which we denote by $u_s(x,t)$,
	plotted in dashed curves.
	Panel C displays realizations of $Y_t$, the process generating the measure under which the 
	solution is calculated.
	The analytical control policy at $t = T$ is given by 
	$u(t) = -\frac{1}{2}[\delta(x - 1) - \delta(x + 1)]$. 
		}
	\label{fig:pimc-example}
\end{figure}

Though nonlinear, this HJB equation can be transformed into a backward Kolmogorov equation (BKE)
through a change of variables.
The BKE can be
solved using path integral methods \cite{kappen2005path}. 
Setting $V(x,t) = -\eta \log \pphi(x,t)$,
substituting in Eq.\ \ref{eq:commit-oc},
and performing the differentiation,
we are able to remove the
nonlinearity if and only if $\frac{\eta^2}{4}\frac{1}{\pphi^2}\left( \frac{\partial \pphi}{\partial x} \right)^2
=
\frac{\sigma^2\eta}{2}\frac{1}{\pphi^2}\left( \frac{\partial \pphi}{\partial x} \right)^2$.
Setting $\eta = 2 \sigma^2$ satisfies this condition.
Performing the change of variables,
Eq.\ \ref{eq:commit-oc} is now linear and has a time-dependent drift and sink term,
\begin{equation}\label{eq:commit-bke}
	\begin{aligned}
		\frac{\partial \pphi}{\partial t} &= \frac{\lambda}{2\sigma^2}v(t)^2\pphi(x,t)
		- v(t)\frac{\partial \pphi}{\partial x} - \frac{\sigma^2}{2}\frac{\partial^2\pphi}{\partial x},\\
		\pphi(x,T) &= \exp \left\{ -\frac{1}{2\sigma^2}\Phi(x) \right\}.
	\end{aligned}
\end{equation}
Application of the Feynman-Kac formula gives the solution to Eq.\ \ref{eq:commit-bke} as 
\cite{kac1949distributions} 
\begin{equation}
	\begin{aligned}
		\pphi(x,t) &= \exp\left\{ - \frac{\lambda}{2\sigma^2}\int_t^Tv(t')^2\ \dee t' \right\}\times\\
		&\quad E_{Y_t}\left\{  
	\exp\left[ -\frac{1}{2\sigma^2}\Phi(Y_T)  \right]
	\bigg| Y_t = x 
	\right\},
	\end{aligned}
\end{equation}
where $Y_t$ is defined by 
\begin{equation}\label{eq:abm-const-control}
	\dee Y_t = v(t)\ \dee t + \sigma \dee \wiener_t,\ Y_0 = x.
\end{equation}
Using this formalism, we apply path integral control to estimate the value function for arbitrary $v(t)$.
Fig.\  \ref{fig:pimc-example} displays example path integral solutions to Eq.\ \ref{eq:commit-oc} when 
player $\lnot i$ credibly commits to playing $v(t) = t^2$ for the duration of the game and 
player $i$'s final cost function takes the form $\Phi(x) = \Theta(|x| - 1) - \Theta( 1 - |x|)$.
In this figure we display the approximate value functions $V(x,t)$
along with their corresponding approximate control
policies $u(x,t) = -\frac{1}{2}\frac{\partial V(x,t)}{\partial x}$.
We calculated these approximations on a grid of $N_x = 500$ linearly spaced $x_n \in [-2, 2]$.
Since the approximated $u(x_n,t)$ are noisy stochastic functions we also plot smoothed versions of them
in Fig.\ \ref{fig:pimc-example}.
These smoothed versions are defined by 
\begin{equation}
	u_s(x_n,t) = \sum_{n'=-k}^k u(x_{n + n'}, t).
\end{equation}
We set $k = 7$ and hence $u_s(x_n, t)$ are $2k + 1 = 15$-step moving averages of the more noisy $u(x_n, t)$.
As $t \rightarrow T$, $u_s(x_n, t)$ approaches the analytical solution of the control problem at $t = T$, 
which is given by $u(x, T) = -\frac{1}{2}[\delta(x - 1) - \delta(x + 1)]$.
\medskip

\noindent
In the further restricted case where there is a credible commitment by one party to play 
a constant control policy $v(t) = v$, we can derive further analytical results.
Under this assumption, the probability law corresponding with Eq.\ \ref{eq:abm-const-control} is given by 
\begin{equation}
	u(y,t) = 
	\frac{1}{\sqrt{2\pi \sigma^2t}}\exp\left\{ \frac{1}{2\sigma^2 t}[(y - x) - vt]^2 \right\},
\end{equation}
so that the (exponentially-transformed) value function reads
\begin{align}\label{eq:exp-value-function}
	&\pphi(x,t) = E_{u(y,T - t)}\left[ 
		\exp\left\{ - \frac{\lambda v^2}{2\sigma^2}(T - t)\Phi(Y_T) \right\}
	\right]\\
	&\quad= \frac{ \exp\left\{ - \frac{\lambda v^2}{2\sigma^2}(T - t) \right\} }
	{\sqrt{2\pi \sigma^2 (T - t)}}
	\times\\
	&\int\displaylimits_{-\infty}^{\infty}
	\exp\left\{-\frac{1}{2\sigma^2}\left[ \Phi(y) 
	+ \frac{((y - x) - v(T - t))^2}{T - t} \right]  \right\} \dee y \notag.
\end{align}
This integral can be evaluated exactly for many $\Phi(y)$ and, for many other final conditions, 
it can be approximated using the method of 
Laplace.
When $t \rightarrow T$  so that the denominator of the argument of the exponential in Eq.\ \ref{eq:exp-value-function}
approaches zero,
Laplace's approximation to the integral reads
\begin{equation}\label{eq:laplace-approx-integral}
\begin{aligned}	
	&\int\displaylimits_{-\infty}^{\infty}
		\exp\left\{-\frac{1}{2\sigma^2}\left[ \Phi(y) 
		+ \frac{((y - x) - v(T - t))^2}{T - t} \right]  \right\} \dee y\\
		&\quad\approx \sqrt{2 \pi \sigma^2 (T -t)}
		\exp\left\{ -\frac{1}{2\sigma^2}\Phi(x +(T - t)v) \right\}.
\end{aligned}
\end{equation}
Inverting the transformation $\varphi$, the value function is approximated by 
\begin{equation}\label{eq:laplace-value}
	V(x,t) = \lambda v^2 (T - t) + \Phi(x + (T - t)v),
\end{equation}
and the control policy by 
\begin{equation}
	u(x, t) = -\frac{1}{2}\Phi'(x + (T - t)v).
\end{equation}
\begin{figure}
\centering
	\includegraphics[width=\columnwidth]{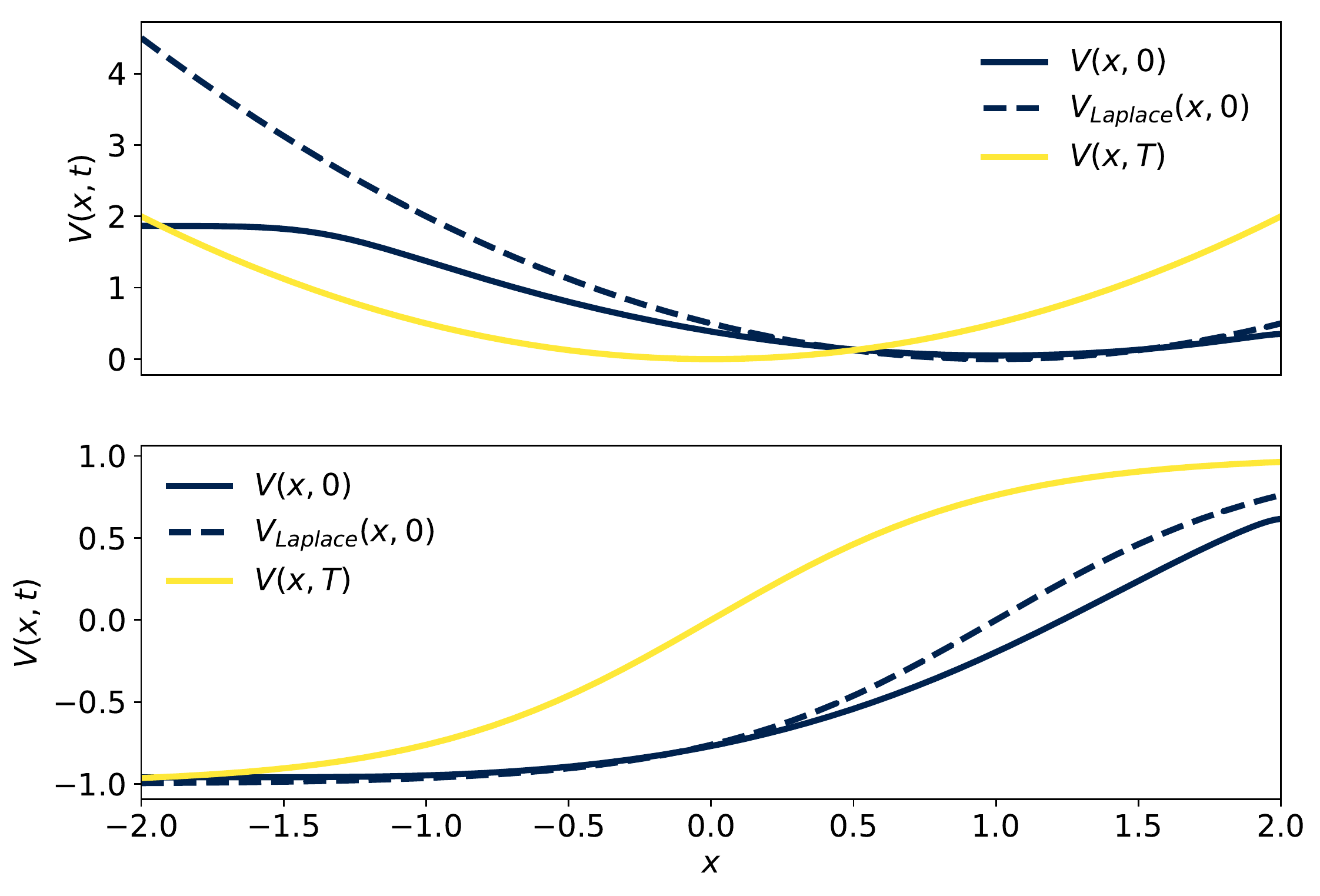}
	\caption{When player $\lnot i$ commits to playing a constant strategy profile $v(t) = v$ for a
	fixed interval of time, 
	an analytic approximate form for player $i$'s value function $V(x,t)$ is given by 
	$V(x,t) \approx \lambda v^2 (T - t) + \Phi(x + (T - t)v)$. 
	We show the numerically-determined value functions 
	at time $t=0$ in solid black curves,
	We display the Laplace approximations at $t=0$ in dashed black curves. 
	The lighter-hue curves are the value functions at the final time $T$, i.e., the 
	respective final conditions.
	The top panel demonstrates results for the final condition $\Phi(x) = \frac{1}{2}x^2$, 
	while the bottom panel has $\Phi(x) = \tanh(x)$.
	}
	\label{fig:laplace-value-approximation}
\end{figure}
We display the results of approximating the value function with 
Eq.\ \ref{eq:laplace-value} at $t = 0$ in Fig.\  \ref{fig:laplace-value-approximation},
along with the actual numerically-determined value function at both $t = 0$ and, for reference, $t = T$.

\subsubsection{Dependence on a free parameter}
The Laplace-approximated value function may depend on a free parameter $a$ that can be used as a
``control knob'' to adjust the approximation.
For example, player $i$ might use $a$ to tune the sensitivity of the approximation to the 
electoral process's distance from a dead-heat.
Ideally, the approximated control policy
should have similar scaling and asymptotic properties as the true control policy.
Solving an optimization problem for optimal values of $a$ is
be one approach to satisfying this desideratum.
\medskip

\noindent
As a case study we consider the behavior of the Laplace-approximated value function 
$V^{(a)}(x,t)$ and its corresponding control policy $u^{(a)}(x,t)$ when we set 
$\Phi^{(a)}(x) = \tanh(ax)$.
We consider this specific example because $\Phi^{(a)}(x) \rightarrow \Theta(x) - \Theta(-x)$
as $a \rightarrow +\infty$.
This limit can be the source of complicated behavior in 
a variety of fields such as piecewise-smooth dynamical systems
(both deterministic and stochastic) \cite{chen2013weak,leifeld2015persistence},
Coulombic friction \cite{kanazawa2015asymptotic}, 
and evolutionary biology \cite{piltz2018inferring}.
\medskip

\begin{figure*}
\centering
	\includegraphics[width=\textwidth]{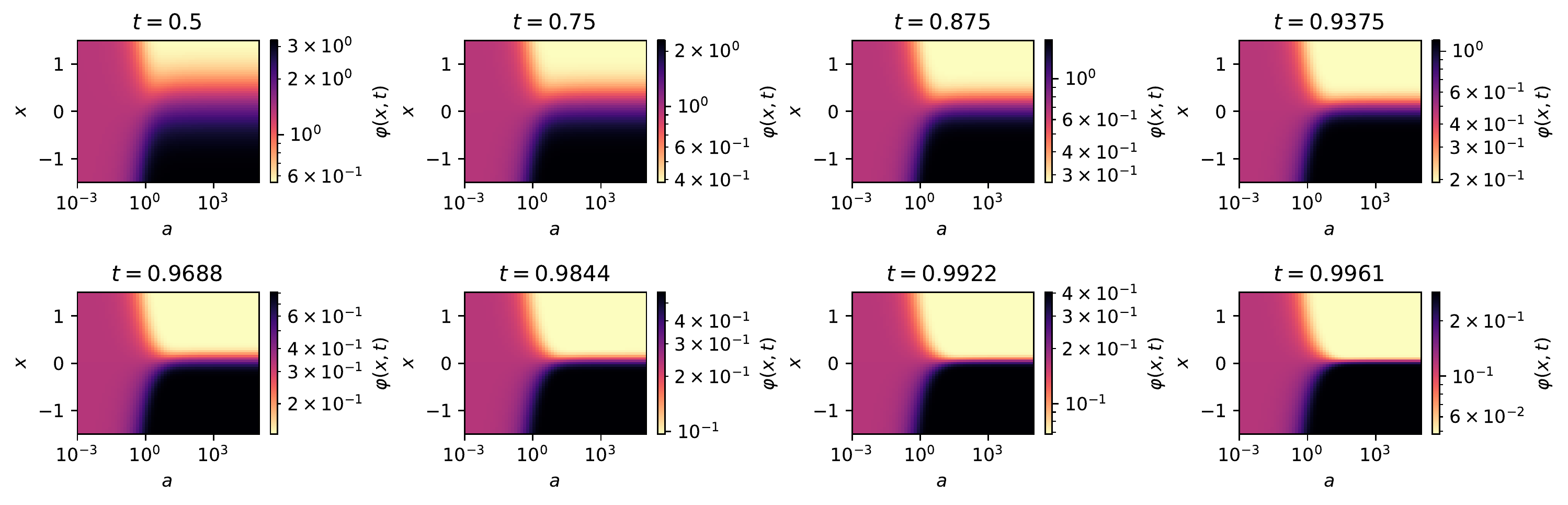}
	\caption{
		If player $\lnot i$ credibly commits to a strategy of playing a constant strategy with value equal to $v$ 
		for the entire duration of the game, 
		player $i$'s (exponentially-transformed) value function $\pphi(x,t)$ has an integral representation given by 
		Eq.\ \ref{eq:exp-value-function}.
		We display dynamics of $\pphi(x,t)$ with the final condition set to $\Phi(x) = \tanh( ax )$.
		We calculated this value function for
		$x \in \left( -\frac{3}{2}, \frac{3}{2} \right)$
		and logarithmically equally-spaced values of $a \in [10^{-3}, 10^5]$.
		For $a < 10^{-1}$, the value function is nearly constant.
		When $a > 10^1$, $\frac{\partial}{\partial x}\pphi(x,t)$ increases in magnitude near $x = 0$
		as $t \rightarrow T$.
		}
	\label{fig:a-x-param-sweep}
\end{figure*}

\noindent
Fig.\ \ref{fig:a-x-param-sweep} displays player $i$'s exponentially-transformed value function 
Eq.\ \ref{eq:exp-value-function}
with final condition $\Phi_i(x) = \tanh(ax)$.
Here player $\lnot i$ credibly commits to playing a constant strategy of $v = 0.01$.
As $t \rightarrow T$, larger values of $a$ lead to a sharp boundary betweeen 
regions of the state space that are costly for player $i$ (positive values of $x$) and those that are 
less costly (negative values of $x$).
This behavior is qualitatively similar to behavior arising from the final condition $\Phi_i(x) = 
\Theta(x) - \Theta(-x)$.
However, we will show that
that there are significant scaling differences in the control policies
resulting from using $\Phi(x)$ versus $\Phi^{(a)}(x)$.
From Eqs.\ \ref{eq:laplace-approx-integral} and \ref{eq:laplace-value},
we approximate the value function by
\begin{equation}
	V^{(a)}(x,t) \approx \lambda v^2(T-t) + \tanh\left\{a[x+v(T-t)]\right\},
\end{equation}
and hence the control policy is approximately 
\begin{equation}\label{eq:approximate-heaviside-control}
	u^{(a)}(x,t) \approx -\frac{a}{2}\sech^2\left\{a[x+v(T-t)]\right\},
\end{equation}
with both expansions increasingly accurate as $t \rightarrow T$.
When $\Phi(x) = \Theta(x) - \Theta(-x)$, we can compute the value function analytically:
\begin{widetext}
\begin{equation}
	\begin{aligned}
		&\frac{1}{\sqrt{2\sigma^2(T-t)}}\int\displaylimits_{-\infty}^{\infty}
		\exp\left\{ -\frac{1}{2\sigma^2}\left[ \Theta(y) - \Theta(-y)
		+ \frac{((y - x) - v(T - t))^2}{T - t}  \right] \right\} \dee y\\
		&\qquad= 
		\cosh\left( \frac{1}{2\sigma^2} \right) 
		+ \sinh\left( \frac{1}{2\sigma^2}  \right) \erf\left( -\frac{x + v(T-t)}{\sqrt{2\sigma^2(T-t)}} \right),
	\end{aligned}
\end{equation}
whereupon we find that
	\begin{equation}\label{eq:exact-heaviside}
		V(x,t) = \lambda v^2(T-t) -2\sigma^2 \log 
		\left[ \cosh\left( \frac{1}{2\sigma^2} \right) 
		+ \sinh\left( \frac{1}{2\sigma^2}  \right)
		\erf\left( -\frac{x + v(T-t)}{\sqrt{2\sigma^2(T-t)}} \right) \right]
	\end{equation}
\end{widetext}
and
\begin{equation}\label{eq:exact-heaviside-control}
		u(x,t) = - \sqrt{
			\frac{2\sigma^2}{\pi(T-t)}
			}
			\frac{
				\exp\left( \frac{-(x + v(T-t))^2}{2\sigma^2(T-t)} \right)
				}
			{
				\coth\left( \frac{1}{2\sigma^2} \right) 
		+ \erf\left( -\frac{x + v(T-t)}{\sqrt{2\sigma^2(T-t)}} \right) 
		}.
\end{equation}
The approximate control policy $u^{(a)}(x, t)$ and the limiting control policy have similar negative 
``bell-like'' shapes but also differ in important ways.
The true control policy decays as a Gaussian modulated by the asymmetric function $\erf(\cdot)$.
The approximate policy decays logistically and hence more slowly than the true control policy.
While the approximate policy is symmetric, the true policy is asymmetric due to the error function term.
Using the Laplace approximation results in more control being applied to the electoral 
process than is optimal.
This is because the tails of $u^{(a)}(x,t)$ are heavier than those of $u(x,t)$.
\medskip

\begin{figure*}
\centering
	\includegraphics[width=\textwidth]{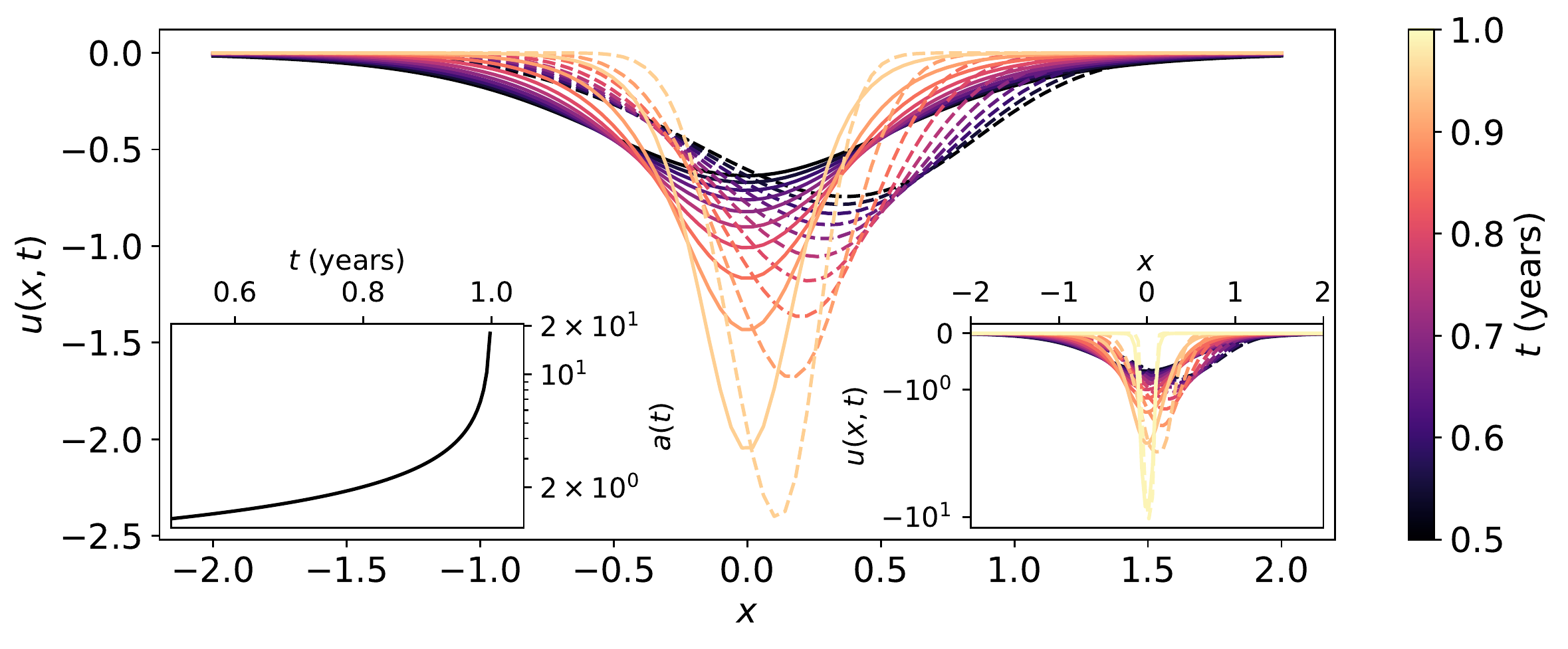}
	\caption{
		The solution to Eq.\ \ref{eq:a-minimization} is a superexponentially-increasing 
		$a(t)$ parameter in the Laplace method-derived value function 
		$V^{(a)}(x,t) = \tanh(ax)$.
		We use this value function as an approximation to the exact value function given in 
		Eq.\ \ref{eq:exact-heaviside}.
		Dashed curves indicate $u(x,t)$, while solid curves indicate $u^{(a)}(x,t)$.
		The lower-right inset axis displays the same data as the main axis and also includes $u(x,t)$ and 
		$u^{(a)}(x,t)$ at the last simulation timestep, $t = 0.9975$, to demonstrate the increasing 
		accuracy of the approximation as $t \rightarrow T$.
		The lower-left inset displays the optimal $a(t)$.
		}
	\label{fig:optimal_a}
\end{figure*}

\noindent
We can maximize the similarity between $u^{(a)}(x,t)$ and $u(x,t)$ by letting the free parameter $a$ be a function
of $t$ and solving the functional minimization problem 
\begin{equation}\label{eq:a-minimization}
	\min_{a(t)} \int\displaylimits_{t}^T 
	\int\displaylimits_{-\infty}^{\infty} [ u^{(a(t))}(x,t) - u(x,t)]^2\ \dee x\ \dee t.
\end{equation}
A stationary point of this problem is given by an $a(t)$ that solves
\begin{equation}\label{eq:integral-zero}
\int\displaylimits_{-\infty}^{\infty}
	[u^{(a(t))}(x,t) - u(x,t)]\frac{\partial u^{(a(t))}(x,t)}{\partial a(t)}\ \dee x = 0.
\end{equation}
We are unable to compute this integral analytically upon substituting Eqs.\ 
\ref{eq:approximate-heaviside-control} and \ref{eq:exact-heaviside-control}.
We find the solution to this problem by numerically solving Eq.\ \ref{eq:integral-zero} using the secant method 
for each of 100 linearly-spaced $t \in \left[0.5, 0.9975\right]$.
We display the optimal $a(t)$, along with the corresponding $u^{(a(t))}(x,t)$ and true $u(x,t)$ in
Fig.\ \ref{fig:optimal_a}.
We find that the optimal $a(t)$ grows superexponentially as $t \rightarrow T$ and that the accuracy of the approximation 
increases in this limit.
This is expected given that $u^{(a)}(x,t)$ is derived using the Laplace approximation and it is 
in this limit that the Laplace approximation is valid.
\medskip

\noindent
Even with the assumption of credible committment to a constant control policy $v$,
we can use this theory to approximate the value function in a noncooperative scenario. 
For arbitrary $v(t)$, expansion about $t + \Delta t$ gives $v(t + \Delta t)
 = v(t) + v'(t)\Delta t,
 $
 leading to an approximate value function iteration over a small time increment $\Delta t$, 
 \begin{equation}
	 V(x, t + \Delta t) \approx \lambda v(t)^2 (T - t) + 
	 \Phi(x + (T - t)[v(t) + v'(t)\Delta t]).
 \end{equation}
In application, both of $v(t)$ and $v'(t)$ can be estimated from possibly-noisy data on $t' \in [0,t]$.

\section{Application}\label{sec:application}
An example of election interference operations is the Russian military foreign 
intelligence service (Red team) activity in the 2016 U.S.\ presidential election contest.
Red team attempted to harm 
one candidate's (Hilary Clinton's) chances of winning and aid another candidate (Donald Trump)
\cite{nyt2019}.
Though Russian foreign intelligence had conducted election interference operations in the past at least 
once before, in the Ukrainian elections of 2014 \cite{tanchak2017invisible}, 
the 2015 and 2016 operations were notable in that Red team operatives used the microblogging website Twitter in 
an attempt to influence the election outcome. 
When this attack vector was discovered, Twitter shut down accounts associated with Red team activity and all 
data associated with those accounts was collected and analyzed \cite{boatwright2018troll,
badawy2018analyzing,boyd2018characterizing}.
There has been analysis of the qualitative and 
statistical effects of these and other election attack vectors (e.g.,
Facebook advertisement purchases) on election polling and the outcome of the election 
\cite{ruck2019internet} and on the detection of election 
influence campaigns more generally \cite{mesnards2018detecting,im2019still}. 
However, to the best of our knowledge, there exists no publicly-available effort to reverse-engineer
the quantitative nature of the control policies used by Russian military intelligence and 
by U.S.\ domestic and foreign intelligence agencies.
\medskip

\begin{figure}[!htp]
\centering
	\includegraphics[width=\columnwidth]{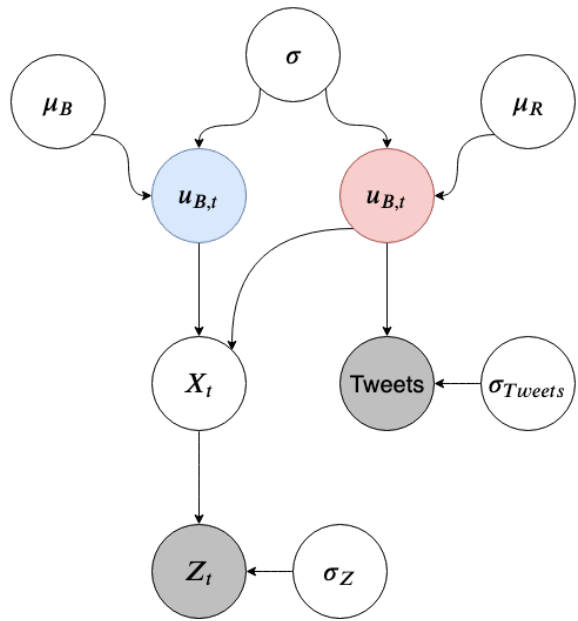}
	\caption{
	We approximate the time series components of the analytical model defined in Sec.\ \ref{sec:neq} 
	by a Bayesian structural time series (BSTS) model.
	We subsequently confront the BSTS model with 2016
	U.S. presidential election data.
	Observed random variables are denoted by gray-shaded nodes, while latent random variables
	are represented by unshaded nodes or red ($u_{\rr,t}$) and blue ($u_{\bb,t}$) nodes.
	We observe a noisy election poll, denoted by $Z_t$, and a time series of tweets associated with Russian
	military intelligence, denoted by $\text{Tweets}$.
	Our objective in this modeling stage 
	is to infer the latent electoral process, denoted by $X_t$, and the latent control 
	policies.
	}
	\label{fig:red-blue-dag}
\end{figure}

\noindent
We first
fit a discrete-time formulation
of the model described in Sec.\ \ref{sec:neq}.
We then compare it to theoretical predictions by
finding values of free parameters in the theoretical model that best describe the observed data and inferred latent controls.
We are faced with two distinct
sources of uncertainty in this procedure.
First, we cannot observe either Red's or Blue's control policy directly 
because foreign and domestic intelligence agencies shroud their activities in secrecy. 
Second, each player's final time payoff structure is also secret and unknown to us. 
To partially circumvent these issues, we construct a two-stage model. 
The first stage is a Bayesian structural time series model, depicted graphically in Fig. \ref{fig:red-blue-dag},
through which we are able to infer distributions of 
discretized analogues of $u_{\rr}(t)$, $u_{\bb}(t)$, and $x(t)$.
Once we have inferred these distributions,
we minimize a loss function that compares the means of these distributions 
to the means of distributions produced by the model described in Sec.\ \ref{sec:neq}. 
\medskip

\noindent
We make the simplifying assumptions about the format of the election that we stated in Sec.\ \ref{sec:intro}
when constructing the discrete-time election model.
Namely, we assume that only two candidates
contest the election and that the election process is modeled by a simple ``candidate A versus candidate B'' poll.
Though there are methods for forecasting elections that make fewer and less restrictive assumptions than these, 
such as compartmental infection models 
\cite{volkening2018forecasting}, 
prediction markets \cite{rothschild2009forecasting},
and more sophisticated Bayesian models \cite{linzer2013dynamic,wang2015forecasting},
we construct our statistical model to mimic the underlying election model of
Sec.\ \ref{sec:neq}.
We do this to test the ability of this underlying theoretical model
to reproduce inferred control and observed election dynamics.
\medskip

\noindent
We can observe neither the Red $u_{\rr}(t)$ nor Blue $u_{\bb}(t)$ control policies.
However, we are able to observe a proxy 
for $u_{\rr}$: the number of tweets sent by Russian military intelligence-associated accounts in the year leading up to the 
2016 election
\footnote{
	Data can be downloaded at 
\www{https://github.com/fivethirtyeight/russian-troll-tweets/}
}. 
This dataset contains a total of 2,973,371 tweets from 2,848 unique Twitter handles.
Of these tweets, a total of 1,107,361 occurred in the year immediately preceding the election (08/11/2015 - 
08/11/2016).
We grouped these tweets by day and used the time series 
of total number of tweets on each day as an observable from which we could infer 
$u_{\rr}$.
We restricted the time range of the model to begin at the later of the end dates of the Republican National
Convention (21 July 2016) and Democratic National Convention (28 July 2016).
We did this because the later of these dates, 
28 July 2016, is the day on which the race was officially between two major party candidates.
Of all Russian military intelligence-associated tweets in 2016,
363,131 occurred during the 102 days beginning on 28 July 2016 and ending the day before Election Day. 
Though the presence of minor party candidates probably played a role in the result of the election,
even the most 
prominent minor parties (Libertarian and Green) received only single-digit support 
\cite{skibba2016pollsters,neville2017constrained}.
We do not model these minor parties and instead consider only the electoral contest
between the two major party candidates.
We used the RealClearPolitics poll aggregation as a proxy for the electoral process itself
\footnote{Data can be downloaded at \url{
	https://www.realclearpolitics.com/epolls/2016/president/us/general_election_trump_vs_clinton_vs_johnson_vs_stein-5952.html}
}, averaging polls that were recorded on the same date
and using the earliest date in the date range of the poll if it 
was conducted over multiple days as the timestamp of that observation.
We weighted all polls equally when averaging.
\medskip

\noindent
Using these two observed random variables,
we fit a Bayesian structural time series model \cite{barber2011bayesian} 
of the form presented in Fig.\  
\ref{fig:red-blue-dag}.
We now describe the structure of the model and explain our choices of priors and likelihood functions.
In the analytical model, we model the latent control policies $u_{\rr}(t)$ and $u_{\bb}(t)$ by 
time- and state-dependent Wiener processes.
To see this, recall that the state equation evolves according to a Wiener process and apply 
It\-{o}'s lemma to the deterministic functions of a random variable
$-\frac{1}{2}\frac{\partial V_{\rr}}{\partial x'}\big|_{x'=x_t}$
and $-\frac{1}{2}\frac{\partial V_{\bb}}{\partial x'}\big|_{x'=x_t}$, which define the control policies.
A discretized version of the Wiener process is a simple Gaussian random walk.
We thus model the latent Red
and Blue control policies by Gaussian random walks:
\begin{align}
	p(u_{R,t}|u_{R,t-1}, \mu_R, \sigma) &= \mathcal{N}(u_{R, t-1} + \mu_R, \sigma^2)\\
	p(u_{B,t}|u_{B,t-1}, \mu_B, \sigma) &= \mathcal{N}(u_{B, t-1} + \mu_B, \sigma^2)
\end{align}
Similarly, we model the latent election process by a discretized version of the state evolution 
equation Eq.\ \ref{eq:state}:
\begin{equation}
	\begin{aligned}
		&p(X_t|X_{t-1}, u_R, u_B) \\
		&\quad =  
		\mathcal{N}(X_{t-1} + u_{B,t-1} - u_{R,t-1}, 1)
	\end{aligned}
\end{equation}
We assume that the latent election model is subject to normal observation error in latent space.
Since we chose a logistic function as the link between the latent and real (on $(0,1)$) election spaces, 
the likelihood for the observed election process is thus given by a Logit-Normal distribution. 
The pdf of this distribution is
\begin{equation}
	p(Z_t|X_t, \sigma_Z) 
	= \sqrt{ \frac{1}{2\pi \sigma_Z^2} }\frac{\exp\left\{  
	-\frac{(\logit(Z_t) - X_t)^2}{2\sigma_Z^2}
		\right\} }{Z_t(1-Z_t)}.
\end{equation}
Though the number of Russian military intelligence tweets
that occur on any given day is obviously a non-negative integer, we chose not to 
model it this way.
A common and simple model for a ``count'' random variable, such as the tweet time series,
is a Poisson distribution with possibly-time-dependent rate parameter
\cite{erlang1909sandsynlighedsregning,rasch1963poisson,cox1972regression,cox1980point}.
This model imposes a strong assumption on the variance of the count distribution (namely, that the 
variance and mean are equal) which does not seem realistic in the context of the tweet data.
Instead of searching for a discrete count distribution that meets some optimality criterion, we instead 
normalized the tweet time series to have zero mean and unit variance, making it a continuous 
random variable rather than a discrete one.
We then shifted the time series so that the 
the new time series was equal to zero on the day during our study 
with the fewest tweets.
We then modeled this time series $\text{Tweets}_t$ by a normal observation likelihood,
\begin{equation}
	p(\text{Tweets}_t|u_{R,t}, \sigma_{\text{Tweets}})
	= \mathcal{N}(u_{R,t}, \sigma_{\text{Tweets}}^2).
\end{equation}
We placed a weakly-informative prior, a Log-Normal distribution, on 
each standard deviation random variable ($\sigma$, $\sigma_Z$, $\sigma_{\text{Tweets}}$),
and zero-centered Normal priors on each mean random variable ($\mu_R, \mu_B$).
This model is high-dimensional, since the latent time series 
$X$, $u_\rr$, and $u_\bb$ are inferred as $T$-dimensional vectors.
In total this model has $3T + 5 = 311$ degrees of freedom.
\medskip

\begin{figure}[!htp]
\centering
	\includegraphics[width=\columnwidth]{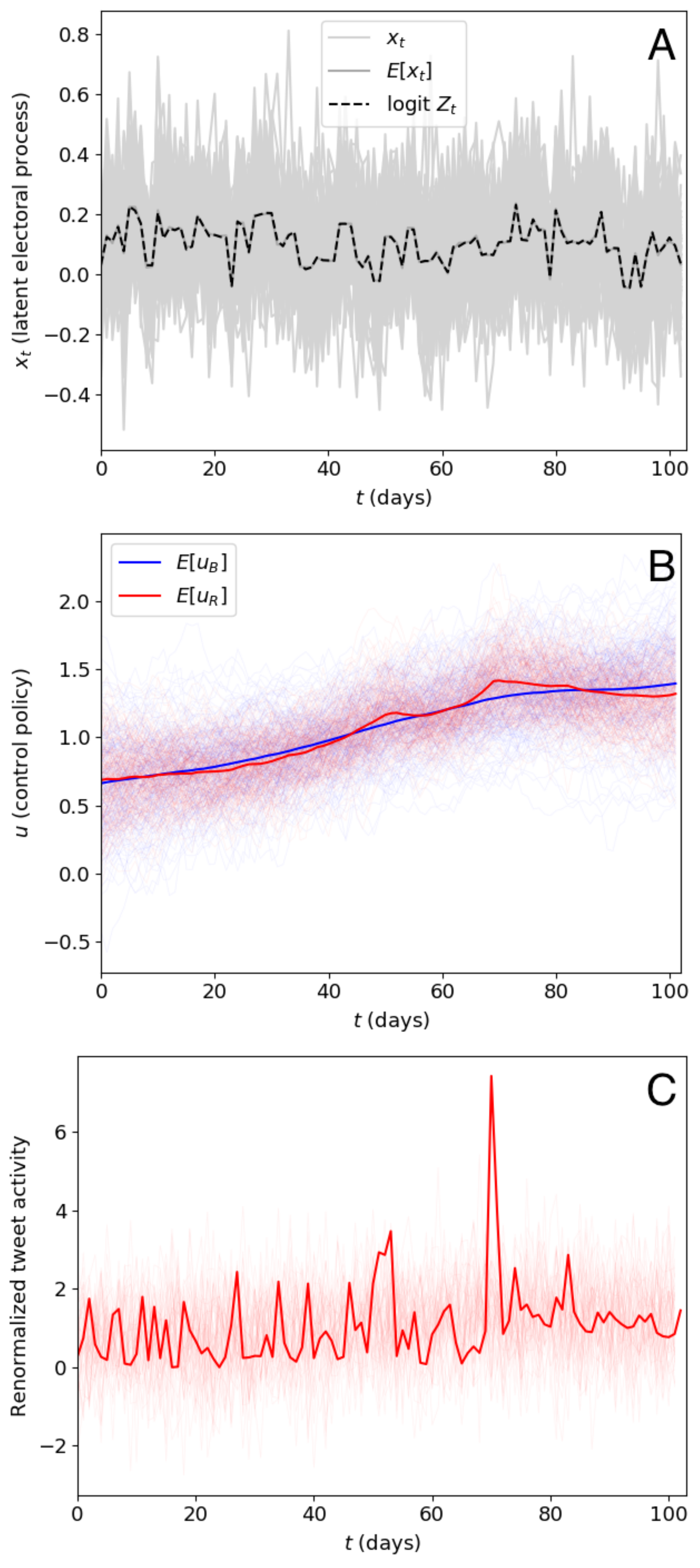}
	\caption{
		Panel A displays the logit of the observed election time series (black curve) $\logit(Z_t)$,
		along with the 
		posterior distribution of the latent electoral process $X_t$.
		Panel B displays the mean latent control policies in thick red and blue curves, along with their 
		posterior distributions.
		Panel C shows the true tweet time series (subject to the normalization described in the main 
		body) along with draws from its posterior predictive distribution.
		The large spike in the tweet time series that is very unlikely under the posterior predictive 
		distribution corresponds to the day (10/06/2016) on which the U.S. federal government officially
		accused Russia of hacking the Democratic National Committee computers.
		}
	\label{fig:bayes-time-series-montage}
\end{figure}

\noindent
We display a graphical representation of this model in Fig.\ \ref{fig:red-blue-dag}.
We fit this model
using the No-U-Turn Sampler algorithm \cite{hoffman2014no}, 
sampling 2000 draws from the model's posterior distribution from each of two independent Markov chains,
We do not include 1000 draws per chain of burn-in in the samples from the posterior.
The sampler appeared to converge well based on graphical consideration (i.e., the ``eye test'')
of draws from the posterior predictive 
distribution of $Z_t$ and $\text{Tweets}_t$, and, more importantly, because 
maximum values of Gelman-Rubin statistics \cite{gelman1992inference}
for all variables satisfied $R_{\max} < 1.01$ except for that of 
$\sigma_Z$, which had $R_{\max} = 1.07646$.
Each of these values is well below the level $R = 1.1$ advocated by Brooks and Gelman \cite{brooks1998general}.
Fig.\  \ref{fig:bayes-time-series-montage} displays draws from the posterior and posterior predictive distribution of
this model. 
Panel A displays draws of $X_t$ from the posterior distribution, along with $E[X_t]$ and $\logit(Z_t)$,
while in panel B we show posterior draws of $u_\rr$ and $u_\bb$, along with $E[u_\rr]$ and $E[u_\bb]$ in thick red and 
blue curves respectively.
In panel C, we display $\text{Tweets}_t$ and draws from its posterior predictive distribution. 
On 10/06/2016, $\text{Tweets}_t$ exhibited a large spike that is very unlikely under the posterior predictive distribution.
This spike likely corresponds with a statement made by the U.S.\ federal government on this date that officially recognized 
the Russian government as culpable for hacking the Democratic National Committee computers.
\medskip

\noindent
After inferring the latent control policies and electoral process, we searched for the parameter values
$\theta = (\lambda_{\rr}, \lambda_{\bb}, \sigma, \Phi_{\rr}, \Phi_{\bb})$ of the theoretical model 
that best explain the observed data and inferred latent variables.
For clarity in reference, we will
refer to the theoretical model as $\modeltwo$ and the Bayesian structural time series model as $\model$.
We use Legendre polynomials to approximate the final conditions $\Phi_{\rr}$ and $\Phi_{\bb}$, 
as discussed in Sec.\ \ref{sec:inference}.
Setting $\Phi_i(x) \approx \sum_{k=0}^K a_{ik}P_k(x)$, $\modeltwo$'s parameter vector is 
$\theta = (\lambda_{\rr}, \lambda_{\bb}, \sigma, a_{0,r},...,a_{K,r}, a_{0,b},...,a_{K,b})$.
 In contrast with $\model$, $\modeltwo$ has relatively few degrees of freedom since the assumption of state and policy 
 co-evolution via solution of coupled partial differential equations
substantially restricts the system's dynamics.
In total, $\modeltwo$ has $2K + 3$ free parameters.
The smaller the value of $K$, the less accurate the approximation to the true final conditions will be.
Conversely, large $K$ could lead to overparameterization of the model and increases the size of the search space.
We thus chose $K = 10$ as a compromise betweeen these two extremes.
With $K = 10$, the model has $2K + 3 = 23$ degrees of freedom.
\medskip

\begin{figure*}
\centering
	\includegraphics[width=\textwidth]{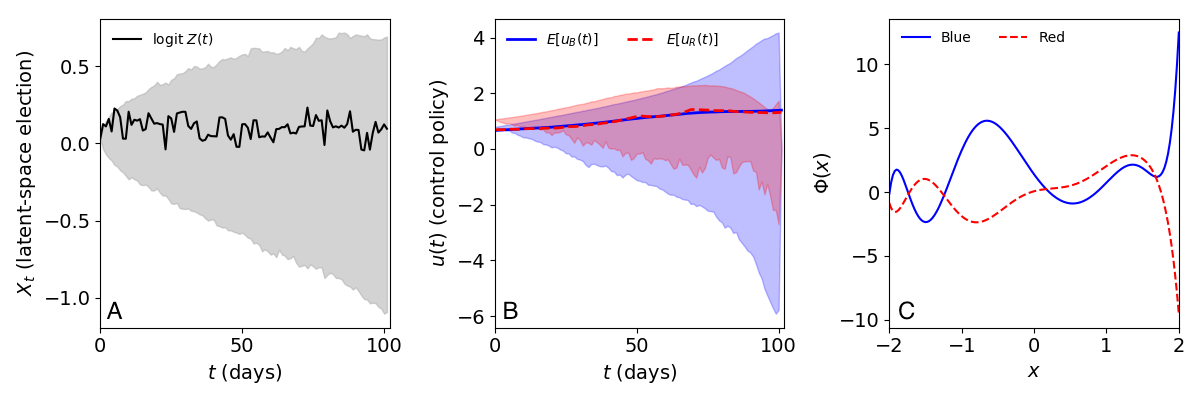}
	\caption{
		We display credible intervals of latent election process $X$ and 
		Red and Blue control policies, $u_\rr$ and $u_\bb$ 
		generated using optimal $\theta = 
		(\lambda_{\rr}, \lambda_{\bb}, \sigma, a_{0,r},...,a_{K,r}, a_{0,b},...,a_{K,b})$ values.
		We ran the optimization algorithm with the number of terms of the Legendre expansion of 
		$\Phi_\rr$ and $\Phi_\bb$ set to $K = 10$ and set the variance regularization in the algorithm to
		$\eta = 0.002$.
		This resulted in fit parameters of 
		$\lambda_\rr = 0.849$, $\lambda_\bb = 0.727$, and $\sigma = 1.509$.
		Panel A displays draws from the latent electoral process under 
		$\modeltwo$, along with $\logit(Z_t)$, the 
		logit-transformed real polling popularity process.
		Panel B displays draws from the distributions of $\hat{u}_\rr$ and $\hat{u}_\bb$ under
		$\modeltwo$, while panel C displays the inferred final conditions $\Phi_\rr(x)$ and 
		$\Phi_\bb(x)$.
		}
	\label{fig:bayes-timeseries-abc}
\end{figure*}

\noindent
The theoretical model $\modeltwo$ can be viewed as a generative probabilitistic function. 
To find optimal parameter values,
we generate $(\hat{u}_\rr, \hat{u}_\bb, \hat{X})$ from $\modeltwo$
and minimize a loss function of these generated values and the values inferred by $\model$.
We defined this loss function as
\begin{equation}\label{eq:loss-fn}
	\begin{aligned}
		L(\theta|\modeltwo) &= \sum_{(y, \hat{y})} \left[
			\left\lVert \mu_y - \mu_{\hat{y}} \right\rVert_2^2 + \eta \sigma_{\hat{y}}
			\right],
	\end{aligned}
\end{equation}
where $y \in \{u_\rr, u_\bb, X\}$ and $\hat{y} \in \{\hat{u}_\rr, \hat{u}_\bb, \hat{X}\}$.
We have defined the mean and standard deviation under the corresponding distribution by $\mu$ and $\sigma$ respectively. 
The $\ell_2$ terms in Eq.\ \ref{eq:loss-fn} penalize deviation 
by $\modeltwo$ from the mean of $\model$'s inferred posterior distribution.
The standard deviation term in Eq.\ \ref{eq:loss-fn} imposes a penalty on dispersion.
We minimized $L(\theta|\modeltwo)$ using a Gaussian-process Bayesian optimization algorithm.
The details of this algorithm are beyond the scope of this work but are readily found in any review paper on the subject
\cite{brochu2010tutorial,shahriari2015taking,frazier2018tutorial}.
Fig.\ \ref{fig:bayes-timeseries-abc} displays the result of this optimization procedure for $K = 10$ and 
$\eta = 0.002$.
For this set of hyperparameters, we found coupling parameter values of $\lambda_\rr = 0.1432$ and $\lambda_\bb = 1.7847$
and a latent space volatility of $\sigma = 0.7510$.
In Fig.\ \ref{fig:bayes-timeseries-abc},
we use credible intervals to denote ranges into which our estimates of model parameters fall.
\medskip

\noindent
Panel A of Fig.\ \ref{fig:bayes-timeseries-abc} displays $\logit(Z_t)$ in a thick black curve and a middle 80\% (10\% to 
90\%) credible interval of $\hat{X}$ from $\modeltwo$ in grey shading.
The observed $\logit(Z_t)$ is centered in the credible interval of $\hat{X}$ and hence 
has a high probability under $\modeltwo$.
In panel B, we show $E[u_\rr]$ and $E[u_\bb]$ in thick red and blue curves respectively along with 
middle 80\% credible intervals of $\hat{u}_\rr$ and $\hat{u}_\rr$.
The mean paths of the latent red and blue control policies
do not lie in the middle 80\% credible intervals 
for approximately the first two weeks after the end of the Democratic National Convention,
but do lie in these credible intervals for the remainder of the time until the election.
Our model is able to capture the election interference dynamics in the middle range of this timespan but is not 
able to capture the dynamics immediately after the race becomes a two-candidate election.
Though the election does officially become a two-candidate 
contest at that time (notwithstanding our previous comments about third-party candidates),
the effects of the Republican and Democratic primaries may take time to dissipate. 
Our model does not capture the dynamics of noncooperative games in the presence 
of many candidates.
We comment on this finding more in Sec.\ \ref{sec:discussion}.
Finally, we display the inferred final conditions $\Phi_\rr(x)$ and 
$\Phi_\bb(x)$ in panel C of Fig.\ \ref{fig:bayes-timeseries-abc}.

\section{Discussion and conclusion} \label{sec:discussion}
We introduce, analyze, and numerically (analytically in simplified cases) solve a simple
model of noncooperative strategic election interference.
This interference is undertaken by a foreign intelligence service from 
one country (Red) in an election occurring in another country (Blue).
Blue's domestic intelligence service attempts to counter this interference.
Though simple, our model is able to provide qualitative insight into the dynamics of such strategic interactions and 
performs well when fitted to polling and social media data surrounding the 2016 U.S.\ presidential election 
contest.
We find that all-or-nothing attitudes regarding the outcome of the election interference,
even if these attitudes are held by only one player,
result in an arms race of spending on 
interference and counter-interference operations by both players.
We then find analytical solutions to player $i$'s optimal control problem when player $\lnot i$
credibly commits to a strategy $v(t)$.
We detail an analytical value function approximation that can 
be used by player $i$ even when player $\lnot i$ does not commit to a particular strategy 
as long as player $\lnot i$'s current strategy and its time derivative can be estimated.
We demonstrate the applicability of our model to real election interference scenarios by 
analyzing the Russian effort to interfere in the 2016 U.S.\ presidential election through observation of 
Russian troll account posts on the website Twitter. 
Using this data, along with aggregate presidential election polling data, we infer the time series of 
Russian and U.S.\ control policies and find parameters of our model that best explain these inferred control policies. 
We show that, for most of the time under consideration (after the Democratic National Convention and before Election Day),
our model provides a good explanation for the 
inferred variables.
However, our model does not accurately or precisely capture the interference dynamics immediately after the 
race becomes a two-candidate race.
\medskip

\begin{figure*}
\centering
	\includegraphics[width=\textwidth]{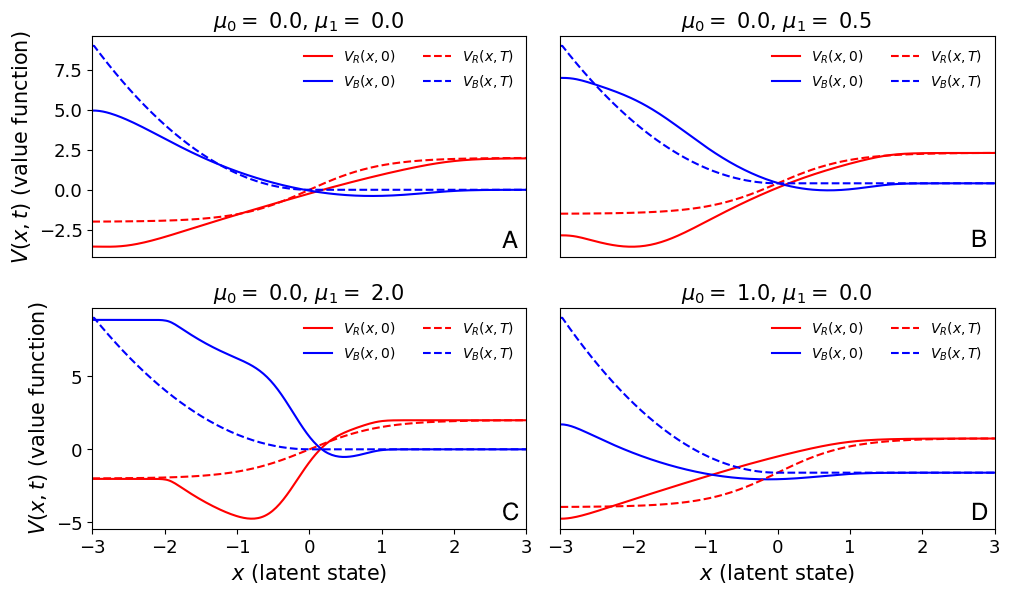}
	\caption{
		We demonstrate the lack of qualitative changes in the value functions given by solutions to
		Eqs.\ \ref{eq:red-gen-value} and \ref{eq:blue-gen-value} when compared to the solutions of Eqs.\
		\ref{eq:red-value} and \ref{eq:blue-value}.
		We plot $V_i(x, 0)$ in solid curves and $V_i(x, T)$ in dashed curves, $i \in \{\rr, \bb\}$.
		We set final conditions here to $\Phi_\rr(x) = \tanh x$ and $\Phi_\bb(x) = \frac{1}{2}x^2 \Theta(-x)$.
		}
		\label{fig:drift-val-func-montage}
\end{figure*}

\begin{figure*}
\centering
	\includegraphics[width=\textwidth]{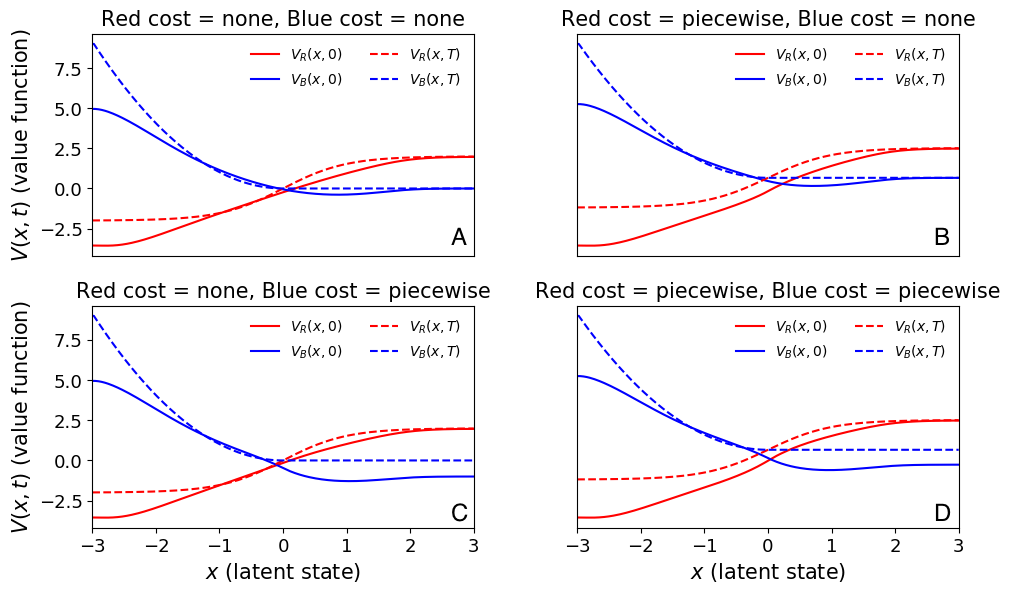}
	\caption{
		We demonstrate the lack of qualitative changes in the value functions given by the solution to
		Eqs.\ \ref{eq:red-gen-value-2} when compared to the solutions of Eqs.\
		\ref{eq:red-value} and \ref{eq:blue-value}.
		We plot $V_i(x, 0)$ in solid curves and $V_i(x, T)$ in dashed curves, $i \in \{\rr, \bb\}$.
		We set final conditions here to $\Phi_\rr(x) = \tanh x$ and $\Phi_\bb(x) = \frac{1}{2}x^2 \Theta(-x)$.
		We also plot solutions for an augmented HJB equation for Blue in panels C and D.
		This HJB equation is identical to Eq.\ \ref{eq:red-gen-value-2} except with $\rr \mapsto \bb$ and the sign 
		of argument of the discontinuous running cost function reversed.
		}
		\label{fig:cost-val-func-montage}
\end{figure*}

\noindent
There are several areas in which our work could be improved. 
While our model is justifiable on the grounds of parsimony and acceptable 
empirical performance on at least one election contest, the kind of assumptions that we make in constructing 
our modeling framework are unrealistic. 
Though a pure random walk model for am election is not 
without serious precedent \cite{taleb2018election}, anextension of this work 
could incorporate non-interference-related
state dynamics as a generalization of Eq.\ \ref{eq:state}.
For example, the state equation could read
\begin{equation}\label{eq:gen-state}
	\dee x = [\mu_0 + \mu_1x + u_{\rr}(t) + u_{\bb}(t)]\dee t + \sigma \dee \wiener.
\end{equation}
This state equation accounts for simple drift in the election results as a candidate endogenously becomes
more or less popular.
It can also account for possible mean-reverting behavior in a hotly-contested race.
Another extension could introduce state-dependent running costs, particularly in the case of the Red player.
Though the action of election interference is nominally intended to cause a particular candidate to win or lose, 
Red could have other objectives as well, such as undermining the Blue citizens' trust in their electoral process. 
Red might gain utility from having a candidate
lead in polls multiple times 
when that candidate would not have otherwise done so, even if the candidate does not actually win the election.
In the context of our model, this is represented by setting Red's cost functional to be 
\begin{equation}\label{eq:gen-red-cost}
	\begin{aligned}
		&E_{u_{\rr}, u_{\bb}, X}\Big\{ \Phi_{\rr}(X_T) + 
	\int_0^T [-\Theta(-X_t)\\
		&\qquad+ u_{\rr}^2(t) - \lambda_{\rr} u_{\bb}^2(t)]\ \dee t \Big\}.
	\end{aligned}
\end{equation}
Both of these modifications are easy to incorporate into the model and do not 
change the qualitative nature of Red and Blue's HJB equations since their effects will simply be 
to introduce an additional drift term (Eq.\ \ref {eq:gen-state}) or 
a continuous, non-differentiable source term (Eq.\ \ref{eq:gen-red-cost}) into the HJB equations 
(Eqs.\ \ref{eq:red-value} and \ref{eq:blue-value}).
That is, the fundamental nature of these equations as 
nonlinear parabolic equations coupled through quadratic terms of self and other-player first spatial 
derivatives remains unchanged as these modifications to the theory do not introduce any new coupling terms.
The solutions to these equations do not demonstrate shock or travelling wave behavior with the addition of the drift or 
source terms,
as we show in Figs.\ \ref{fig:drift-val-func-montage} and \ref{fig:cost-val-func-montage}.
With the modification of Eq.\ \ref{eq:gen-state}, the HJB equations become 
\begin{equation}\label{eq:red-gen-value}
	\begin{aligned}
		&-\frac{\partial V_{\rr}}{\partial t} = 
	(\mu_0 + \mu_1 x)\frac{\partial V_\rr}{\partial x}
	-\frac{1}{4}\left( \frac{\partial V_{\rr}}{\partial x} \right)^2 \\
		&\quad-
	\frac{1}{2}\frac{\partial V_{\rr}}{\partial x}\frac{\partial V_{\bb}}{\partial x}
	- \frac{\lambda_{\rr}}{4}\left(\frac{\partial V_{\bb}}{\partial x}\right)^2
	+ \frac{\sigma^2}{2}\frac{\partial^2 V_{\rr}}{\partial x^2},\\
		&\qquad V_{\rr}(x,T) = \Phi_{\rr}(x)
	\end{aligned}
\end{equation}
and
\begin{equation}\label{eq:blue-gen-value}
	\begin{aligned}
		&-\frac{\partial V_{\bb}}{\partial t} =
	(\mu_0 + \mu_1 x)\frac{\partial V_\bb}{\partial x}
	-\frac{1}{4}\left( \frac{\partial V_{\bb}}{\partial x} \right)^2 \\
		&\quad-
	\frac{1}{2}\frac{\partial V_{\bb}}{\partial x}\frac{\partial V_{\rr}}{\partial x}
	- \frac{\lambda_{\bb}}{4}\left(\frac{\partial V_{\rr}}{\partial x}\right)^2
	+ \frac{\sigma^2}{2}\frac{\partial^2 V_{\bb}}{\partial x^2},\\
		&\qquad V_{\bb}(x,T) = \Phi_{\bb}(x).
	\end{aligned}
\end{equation}
In Fig.\ \ref{fig:drift-val-func-montage} we give examples of
solutions of Eqs.\ \ref{eq:red-gen-value} and \ref{eq:blue-gen-value} at $t = 0$ and $t = T$.
These solutions do not display qualitative changes, such as the formation of shock or travelling waves,
with the inclusion of nonzero drift terms
of the form $\mu_0 + \mu_1 \frac{\partial V_i}{\partial x}$, $i \in \{\rr, \bb\}$.
With the modification of Eq.\ \ref{eq:gen-red-cost},
Red's HJB equation reads
\begin{equation}\label{eq:red-gen-value-2}
	\begin{aligned}
		&-\frac{\partial V_{\rr}}{\partial t} = 
	-\frac{1}{4}\left( \frac{\partial V_{\rr}}{\partial x} \right)^2 -\frac{1}{2}\frac{\partial V_{\rr}}{\partial x}\frac{\partial V_{\bb}}{\partial x}\\
		&- \frac{\lambda_{\rr}}{4}\left(\frac{\partial V_{\bb}}{\partial x}\right)^2
		-\Theta(-x)
	+ \frac{\sigma^2}{2}\frac{\partial^2 V_{\rr}}{\partial x^2},\\
		&\qquad V_{\rr}(x,T) = \Phi_{\rr}(x).
	\end{aligned}
\end{equation}
We plot solutions of Eq.\ \ref{eq:red-gen-value-2} in Fig.\ \ref{fig:cost-val-func-montage}.
These solutions also do not change qualitatively from the solutions to Eqs.\ \ref{eq:red-value} and \ref{eq:blue-value} 
in that there is no shock or travelling wave formation
\footnote{
	Using the HJB integrator \texttt{red\_blue\_pdes.py} included in the code 
	at \www{https://gitlab.com/daviddewhurst/red-blue-game},
	the reader may further investigate the effects of changes in the drift and running cost terms on the 
	qualitative nature of solutions to the coupled HJB system.
}.
A more fundamental qualitative change would be to expand the scope of Red's interference to alter the 
latent volatility of the election process. 
Red's additional objective might be to increase the uncertainty in polling results.
\medskip

\noindent
In addition to theoretical modifications, other work could extend these results to other elections using
similarly fine-grained or more granular data.
This approach is difficult because 
there is very little granular public data on election interference \cite{levin2019partisan}.
We are able to confront our model to data
only because the Russian interference in the 2016 U.S.\ presidential election was well-publicized and 
because the interference took place at least partially through the mechanism of Twitter, which is a public data source.
We were unable to find any other publicly-available 
data at daily (or finer) temporal resolution for any other publicly-acknowledged election interference episode.
\medskip

\noindent
In the case study of the 2016 U.S.\ presidential election,
our theoretical model accurately captured election interference dynamics from approximately two weeks after
the Democratic National Convention until Election Day.
However, it did not capture the dynamics of election interference accurately or precisely during the first 
fortnight of time under study.
We believe this is because, even though the election was then a two-candidate contest,
there were additional election state and interference dynamics that we did not model.
We believe that these dynamics arise because the transition from interfering in many candidates' primary campaigns
to interfering in only one electoral contest is not immediate.
It likely takes time for the foreign intelligence agency to recalibrate their interference strategy.
In addition, the foreign intelligence agency may still expend reseources on influencing other candidates' 
supporters, even though those other candidates had been unsuccessful in their quest for inclusion in the 
general election.
Suppose that there are initially $N_\rr$ ``red candidates'' (candidates that the foreign intelligence service 
would like to win the election) and $N_\bb$ ``blue candidates'' (candidates that the foreign intelligence 
service would not like to win the election).
Then modeling the transition between the conventions and the general election requires collapsing the 
state equation from a $N_\rr + N_\bb - 2$-dimensional stochastic differential equation (SDE) to a one-dimensional 
SDE.
The cost functions of Red and Blue would probably also change during this transition, but we are unsure of how to 
model this change.
Because of the dimensionality reduction in the state equation, the coupled HJB equations would change from being 
PDEs solved in $N_\rr + N_\bb - 2$ spatial dimensions to ones solved in one spatial dimension, as now.
We did not attempt to model these dynamics, but this could be a useful expansion of our model.
\medskip

\noindent
We used two models in our analysis of interference in the 2016 U.S.\ presidential election.
We used a Bayesian structural time series model
to infer the latent random variables $u_\rr$, $u_\bb$, and $X$,
and then used these inferred values to fit parameters of the theoretical model
described in Sec.\ \ref{sec:sgpe}.
While the theoretical model does not have many free parameters ($2K + 3 = 23$ degrees of freedom),
the structural time series model does have many free parameters ($3T + 5 = 311$ degrees of freedom).
The large number of free parameters of the structural time series model 
does not mean that the model is overparameterized.
At each $t$, we observe a number of tweets $\text{Tweets}_t$ and a popularity rating for the candidates $Z_t$.
From these we want to infer the distributions of the random variables $u_{\rr,t}$, $u_{\bb,t}$, and $X_t$.
Since we want to infer the distribution of each of these three random variables at each of the $T$ timesteps, 
this large number of parameters is expressly necessary.
If we observe $M$ identical trials of an election process over $T$ timesteps, 
the ratio of structural time series model parameters to observed datapoints is given by 
\begin{equation}\label{eq:param-ratio}
	R(T,M) = \frac{3T + 5}{2MT} = \frac{3}{2}M^{-1}\left(1 + \frac{5}{3}T^{-1}\right).
\end{equation}
With $T$ held constant, $R(T,M) \rightarrow 0$ as $M$ grows, while with $M$ held constant,
$R(T,M) \rightarrow \frac{3}{2}$ as $T$ grows large.
Since we observe only one draw from the election interference model, $M = 1$ in our case.
However, this approach of inferring each random variable's distribution is 
not tractable when $T$ becomes large since the number of parameters to fit still grows linearly with $T$.
\medskip

\noindent
One way of partially circumventing this problem is to use a variational inference approach combined with amortization 
of the random variables.
Denote the vector of all observed random variables at time $t$ by $y_t$ 
and the vector of all latent random variables at time 
$t$ by $w_t$.
Variational inference replaces the actual posterior distribution with an approximate posterior distribution that has a 
known normalization constant (thereby eliminating the need for MCMC routines to compute this constant)
\cite{hoffman2013stochastic,rezende2015variational,blei2017variational}.
The parameters of the approximate posterior are found through optimization, which is generally much faster 
than Monte Carlo sampling.
If the joint distribution of $y = (y_1,...,y_T)$ and $w = (w_1,...,w_T)$ is given by 
$p(y, w) = \prod_{t=1}^T p(y_t|w_t)p(w_t)$, then the true posterior is given by 
$p(w|y) = p(y, w) / p(y)$.
The approximate (variational) posterior is given by $q_\theta(w) = \prod_{t=1}^T q_{\theta_t}(w_t)$,
where $\theta = (\theta_1,...,\theta_T)$ is the vector of parameters found through optimization and 
$q_{\theta_t}(w_t)$ are the probability distributions for each timestep.
The normalizing constants are known for each $q_{\theta_t}(w_t)$.
\medskip

\noindent
Amortization of the random variables means that, instead of finding the optimal value of the entire
length-$T$ vector $\theta$, we model the approximate posterior as 
$q_{\psi}(w) = \prod_{t=1}^T q(w_t|f_{\psi}(y_t))$ \cite{rezende2015variational,zhang2018advances}.
The vector $\psi$ is the vector of parameters of the (probably nonlinear) function $f_\psi(\cdot)$ and does not 
scale with $T$.
The function $f_\psi(\cdot)$ models the effect of the time-dependent parameters 
$\theta_t$ of the variational posterior.
This amortized variational posterior is also fit using an optimization routine. 
Since the number of parameters of this model does not scale with time, 
Eq.\ \ref{eq:param-ratio} for this model becomes
\begin{equation}
	R_{\text{amort}}(T, M) = \frac{P}{2MT},
\end{equation}
where $P$ is the constant number of parameters (the dimension of $\psi$) in the amortized model.
For fixed $M$, $R_{\text{amort}}(M, T) \rightarrow 0$ as $T$ becomes large.
Another useful extension of our present work would be to reimplement our Bayesian structural time series model
using amortized variational inference.
This would also eliminate the problem of choosing the ``correct'' number of parameters in the Legendre polynomial
approximation that we described in Sec.\ \ref{sec:application}.

\section*{Acknowledgements}
The authors are grateful for financial support from the
Massachusetts Mutual Life Insurance Company and are 
thankful for the truly helpful comments from an 
anonymous reviewer.

\bibliography{red-blue-game}{}
\bibliographystyle{unsrt}

\clearpage
\appendix

\section{Coupling parameter sweeps}
\label{app:lambda-param-sweeps}

We conducted parameter sweeps over different values of the coupling parameters 
$\lambda_\rr$ and $\lambda_\bb$ for multiple combinations of Red and Blue final conditions.
To do this, we integrated Eqs.\ \ref{eq:red-value} and \ref{eq:blue-value}
for $\lambda_\rr, \lambda_\bb \in [0, 3]$ for each of the $3^2 = 9$ combinations 
of the final conditions 
\begin{equation}
\Phi_\rr(x) \in
\left\{ x, 2\tanh x, \Theta(x) - \Theta(-x) \right\}
\end{equation}
and  
\begin{equation}
\Phi_\bb(x) \in \left\{ \frac{1}{2}x^2, \frac{1}{2}x^2\Theta(-x), 2[\Theta(|x| - \Delta) - \Theta(\Delta - |x|)] \right\}.
\end{equation}
We integrated Eqs.\ \ref{eq:red-value} and \ref{eq:blue-value} for $N_t = 8000$ timesteps, setting 
$t_0 = 0$ and $T = 1$ year.
We enforced Neumann boundary conditions on the interval $x \in [-3, 3]$ and set the spatial step size to be 
$\dee x = 0.025$.
After integration, we then drew $1000$ paths $u_\rr^{(n)}$ and $u_\bb^{(n)}$ from the 
resulting probability distribution over control policies.
(This probability distribution is a time-dependent multivariate Gaussian, 
since the control policies are deterministic functions 
of the random variable defined by the Ito stochastic differential equation Eq.\ \ref{eq:state}.)
We then calculated the intertemporal means and standard deviations of these control policies,
which we denote by 
\begin{equation}
	\text{mean}(u_i) = \frac{1}{TN}\sum_{n=1}^N \int_0^T u_i^{(n)}(t)\ \dee t
\end{equation}
and
\begin{equation}
	\text{std}(u_i) =
	\sqrt{\frac{1}{TN}\sum_{n=1}^N\int_0^T \left[u_i^{(n)}(t) - \text{mean}(u_i)\right]^2\ \dee t},
\end{equation}
for $i \in \{\rr, \bb\}$.
We display the mean control policies in Sec.\ \ref{sec:app-means} and 
the standard deviations of the control policies in Sec.\ \ref{sec:app-stds}.

\subsection{Expected value of $u_i(t)$}\label{sec:app-means}
We display the mean paths of draws from the distributions
of control policies generated by solutions to Eqs.\ \ref{eq:red-value} and
\ref{eq:blue-value}.
\input{red_blue_lambda_mean_appendix}  %% generated automatically
\clearpage

\subsection{Standard deviation of $u_i(t)$}\label{sec:app-stds}
We display the standard deviation of draws from the distributions
of control policies generated by solutions to Eqs.\ \ref{eq:red-value} and
\ref{eq:blue-value}.

\input{red_blue_lambda_std_appendix}  %% generated automatically
\clearpage

\end{document}

%% file: red-blue-game.settings.tex
\usepackage{natbib}

\usepackage{color}

\usepackage{graphicx,epsfig,verbatim,enumerate}
\usepackage{amssymb,amsmath,amsthm,amsbsy}
\usepackage{ifthen}
\usepackage{morefloats}

\usepackage{algorithm}
\usepackage{algpseudocode} 

\usepackage{longtable}
\usepackage{booktabs}
\usepackage{hyperref}
\hypersetup{colorlinks=true, linkcolor=blue}

\usepackage{textcomp}
\usepackage{dsfont}

\usepackage{mathtools}
\usepackage{fancyhdr}

\newboolean{twocolswitch}

\newcommand{\eps}{\varepsilon}
\newcommand{\wiener}{\mathcal{W}}
\newcommand{\bigoh}{\mathcal{O}}
\newcommand{\pphi}{\varphi}

\hypersetup{colorlinks=true, linkcolor=blue, citecolor=red}

\newcommand{\sindex}[1]{}
\newcommand{\nindex}[1]{}

\newcommand{\www}[1]{\url{#1}}

\usepackage{lettrine}

\newcommand{\dee}[1]{\textnormal{d}#1}

\newcommand{\rr}{\mathrm{R}}
\newcommand{\bb}{\mathrm{B}}
\newcommand{\sech}{\mathrm{sech}}
\newcommand{\erf}{\mathrm{erf}}
\newcommand{\logit}{\mathrm{logit}}
\newcommand{\model}{\mathcal{M}}
\newcommand{\modeltwo}{\mathcal{Q}}

%% file: red_blue_lambda_mean_appendix.tex
\begin{figure}[!ht]
    \centering
\includegraphics[width=0.9\linewidth]{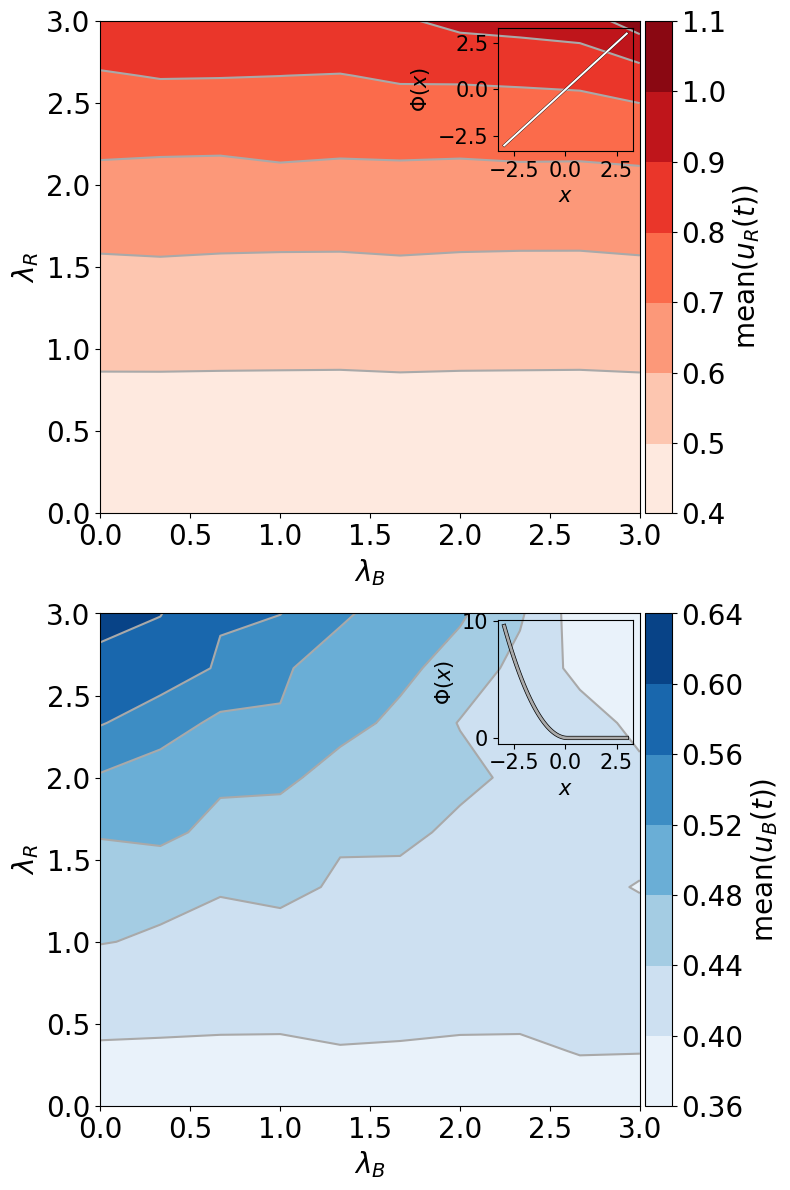}
\caption{Parameter sweep over coupling parameters $\lambda_{\mathrm{R}}, \lambda_{\mathrm{B}}$ with Red final condition $\Phi_{\mathrm{R}}(x) =$ $x$ and Blue final condition $\Phi_{\mathrm{B}}(x) =$ $\frac{1}{2}x^2\Theta(-x)$. Intensity of color corresponds to mean of control policy.}
\end{figure}
\begin{figure}[!ht]
    \centering
\includegraphics[width=0.9\linewidth]{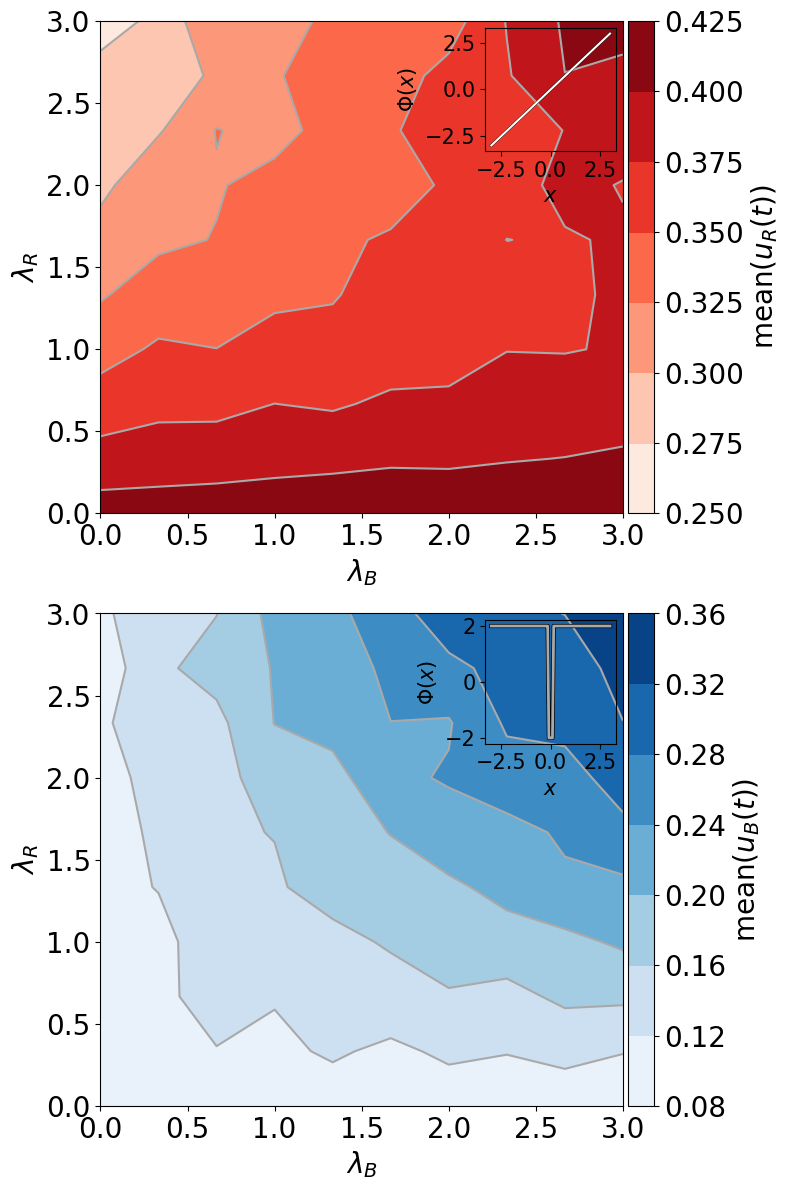}
\caption{Parameter sweep over coupling parameters $\lambda_{\mathrm{R}}, \lambda_{\mathrm{B}}$ with Red final condition $\Phi_{\mathrm{R}}(x) =$ $x$ and Blue final condition $\Phi_{\mathrm{B}}(x) =$ $2[\Theta(|x| - 0.1) - \Theta(0.1 - |x|)]$. Intensity of color corresponds to mean of control policy.}
\end{figure}
\clearpage
\begin{figure}[!ht]
    \centering
\includegraphics[width=0.9\linewidth]{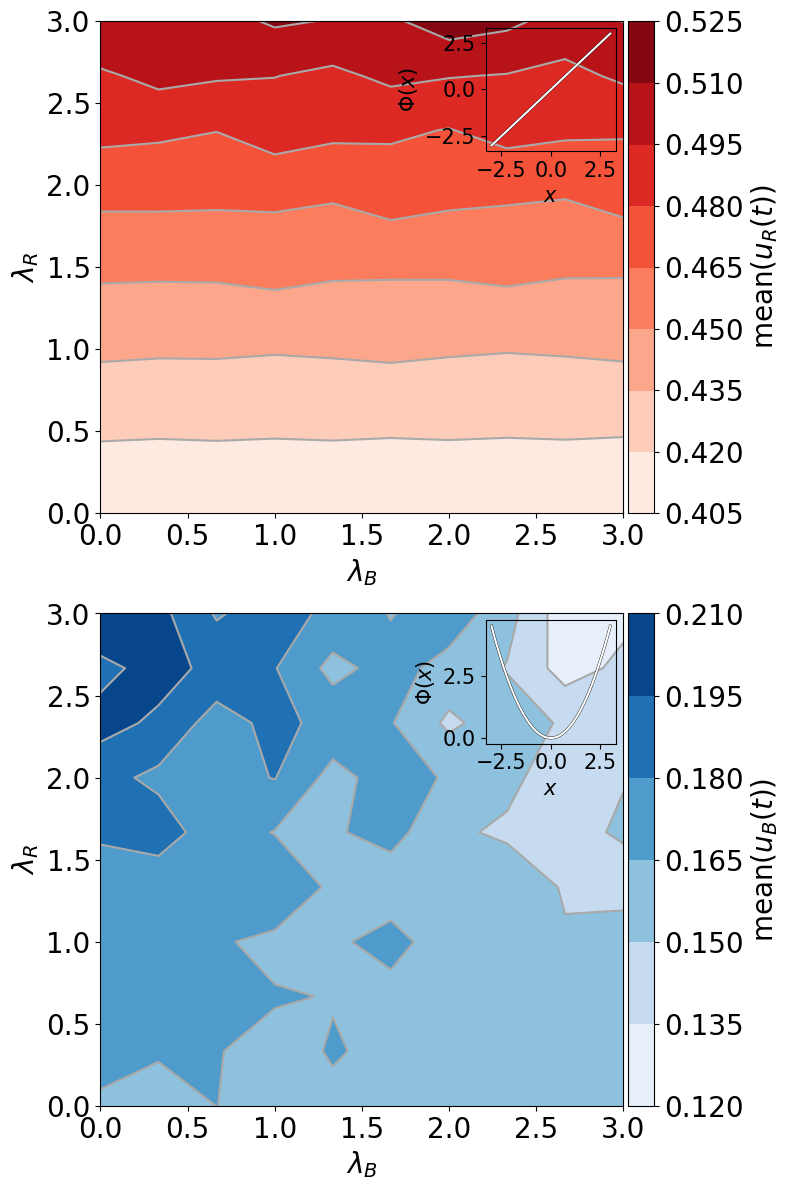}
\caption{Parameter sweep over coupling parameters $\lambda_{\mathrm{R}}, \lambda_{\mathrm{B}}$ with Red final condition $\Phi_{\mathrm{R}}(x) =$ $x$ and Blue final condition $\Phi_{\mathrm{B}}(x) =$ $\frac{1}{2}x^2$. Intensity of color corresponds to mean of control policy.}
\end{figure}
\begin{figure}[!ht]
    \centering
\includegraphics[width=0.9\linewidth]{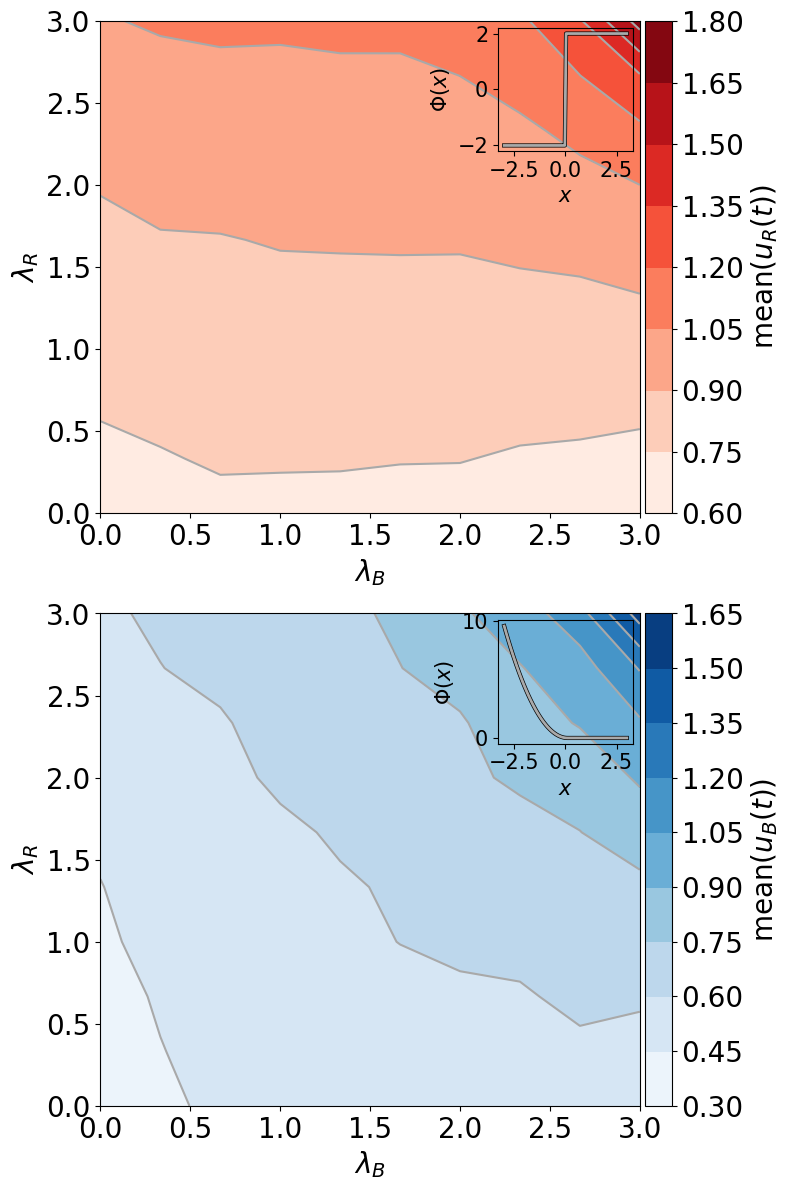}
\caption{Parameter sweep over coupling parameters $\lambda_{\mathrm{R}}, \lambda_{\mathrm{B}}$ with Red final condition $\Phi_{\mathrm{R}}(x) =$ $2[\Theta(x) - \Theta(-x)]$ and Blue final condition $\Phi_{\mathrm{B}}(x) =$ $\frac{1}{2}x^2\Theta(-x)$. Intensity of color corresponds to mean of control policy.}
\end{figure}
\clearpage
\begin{figure}[!ht]
    \centering
\includegraphics[width=0.9\linewidth]{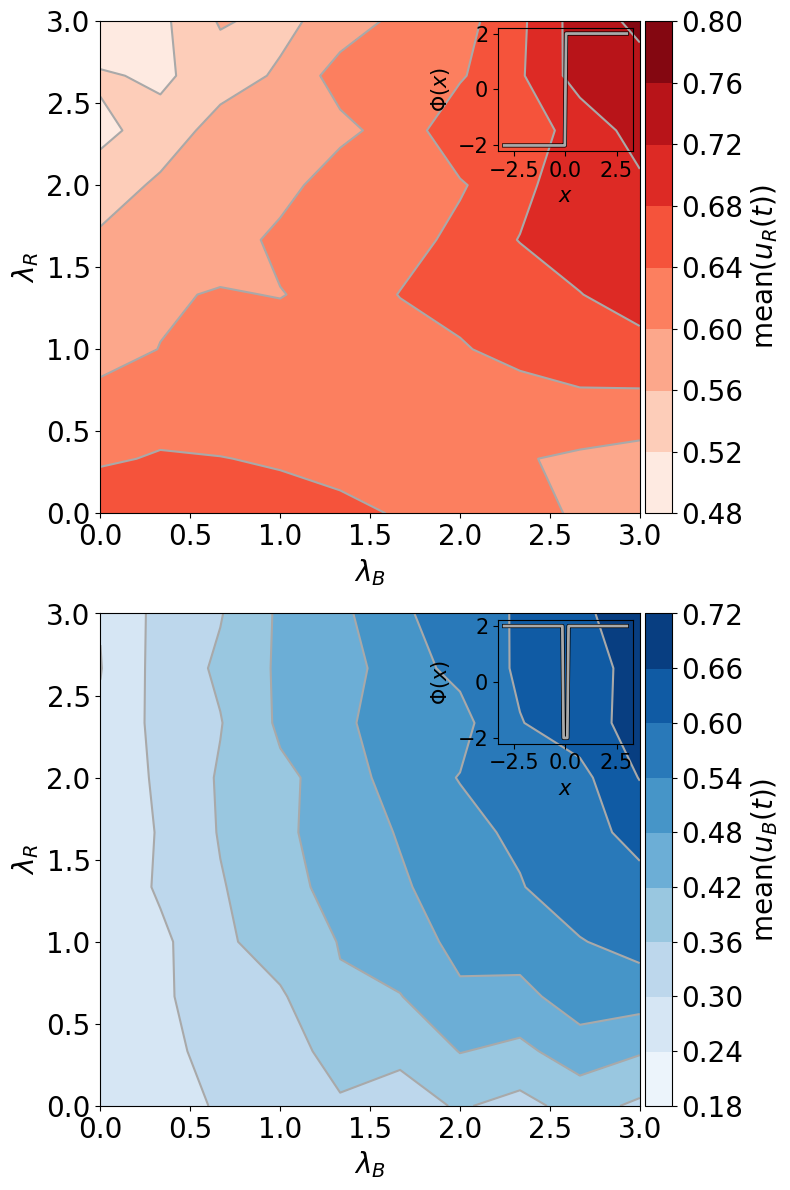}
\caption{Parameter sweep over coupling parameters $\lambda_{\mathrm{R}}, \lambda_{\mathrm{B}}$ with Red final condition $\Phi_{\mathrm{R}}(x) =$ $2[\Theta(x) - \Theta(-x)]$ and Blue final condition $\Phi_{\mathrm{B}}(x) =$ $2[\Theta(|x| - 0.1) - \Theta(0.1 - |x|)]$. Intensity of color corresponds to mean of control policy.}
\end{figure}
\begin{figure}[!ht]
    \centering
\includegraphics[width=0.9\linewidth]{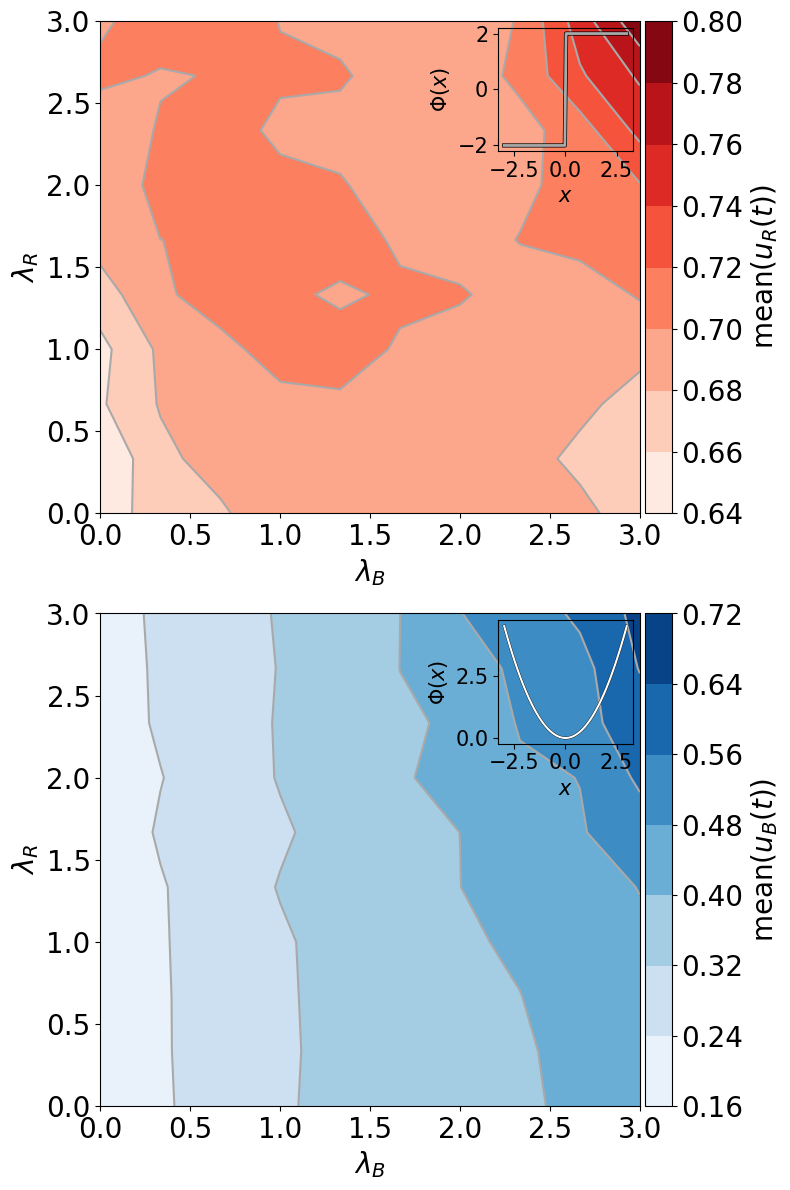}
\caption{Parameter sweep over coupling parameters $\lambda_{\mathrm{R}}, \lambda_{\mathrm{B}}$ with Red final condition $\Phi_{\mathrm{R}}(x) =$ $2[\Theta(x) - \Theta(-x)]$ and Blue final condition $\Phi_{\mathrm{B}}(x) =$ $\frac{1}{2}x^2$. Intensity of color corresponds to mean of control policy.}
\end{figure}
\clearpage
\begin{figure}[!ht]
    \centering
\includegraphics[width=0.9\linewidth]{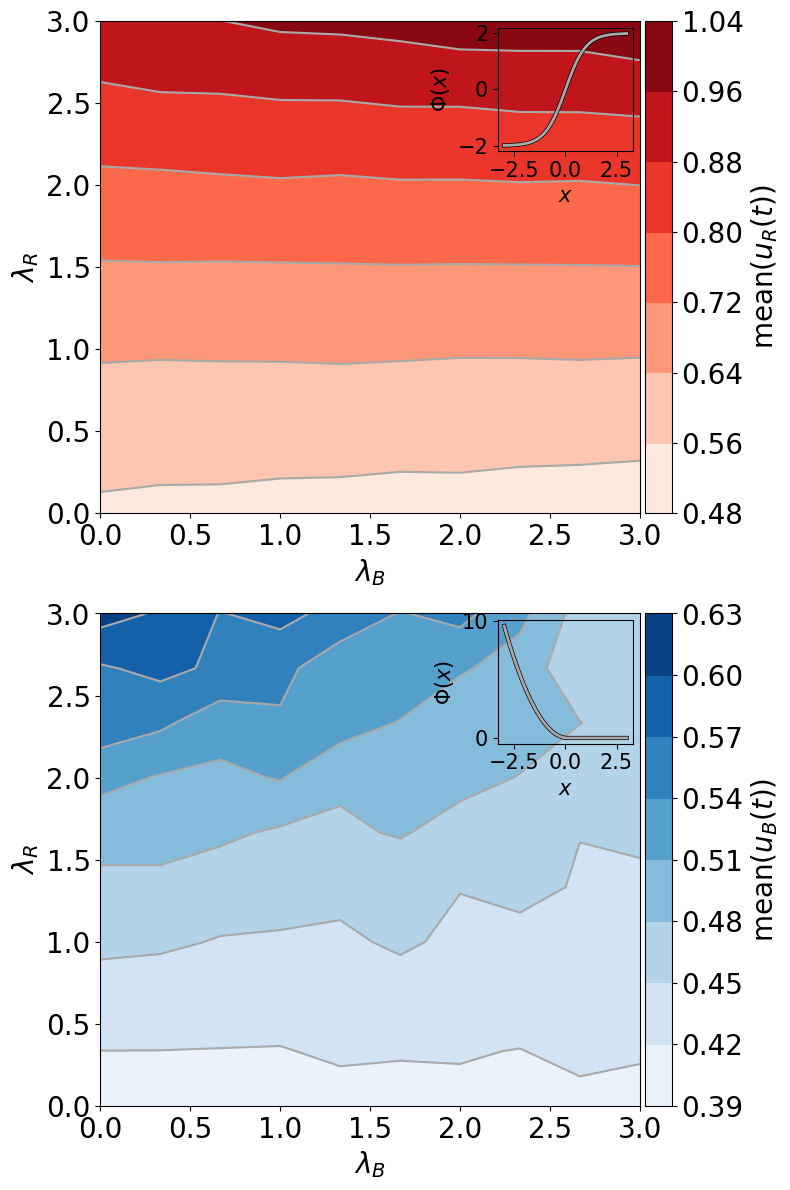}
\caption{Parameter sweep over coupling parameters $\lambda_{\mathrm{R}}, \lambda_{\mathrm{B}}$ with Red final condition $\Phi_{\mathrm{R}}(x) =$ $\tanh(x)$ and Blue final condition $\Phi_{\mathrm{B}}(x) =$ $\frac{1}{2}x^2\Theta(-x)$. Intensity of color corresponds to mean of control policy.}
\end{figure}
\begin{figure}[!ht]
    \centering
\includegraphics[width=0.9\linewidth]{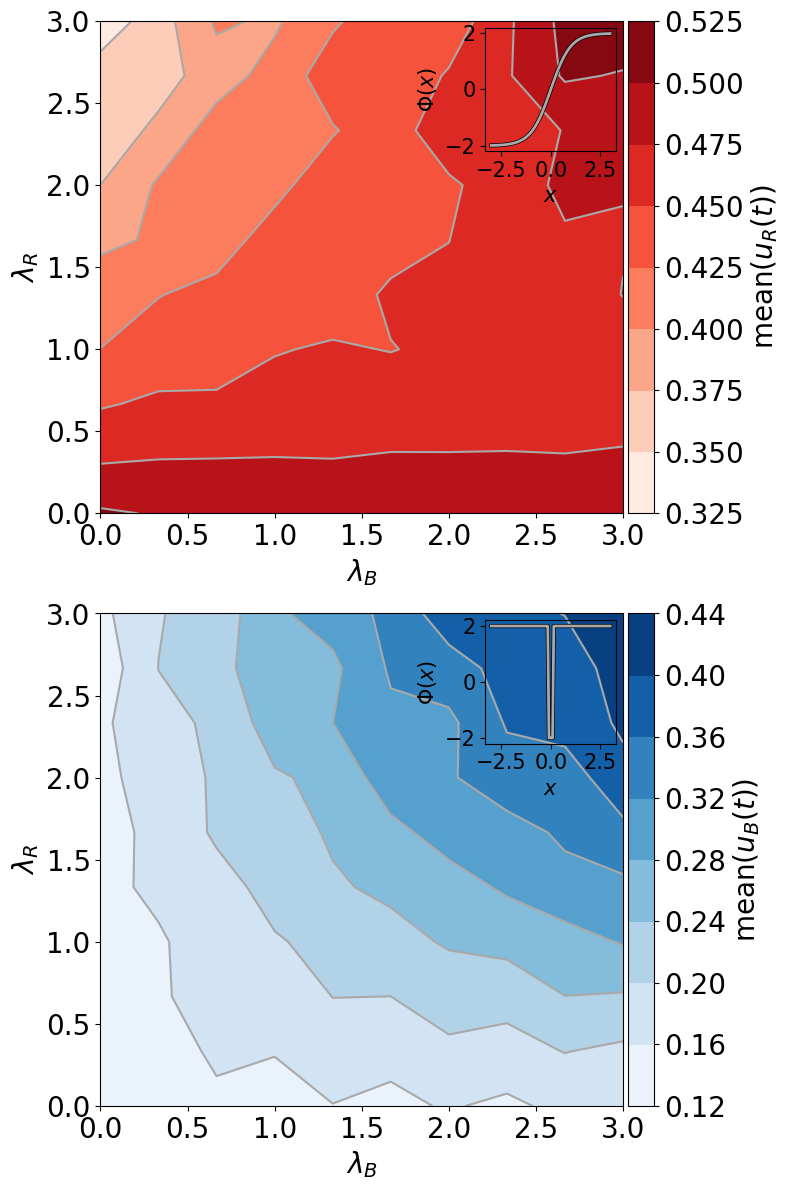}
\caption{Parameter sweep over coupling parameters $\lambda_{\mathrm{R}}, \lambda_{\mathrm{B}}$ with Red final condition $\Phi_{\mathrm{R}}(x) =$ $\tanh(x)$ and Blue final condition $\Phi_{\mathrm{B}}(x) =$ $2[\Theta(|x| - 0.1) - \Theta(0.1 - |x|)]$. Intensity of color corresponds to mean of control policy.}
\end{figure}
\clearpage
\begin{figure}[!ht]
    \centering
\includegraphics[width=0.9\linewidth]{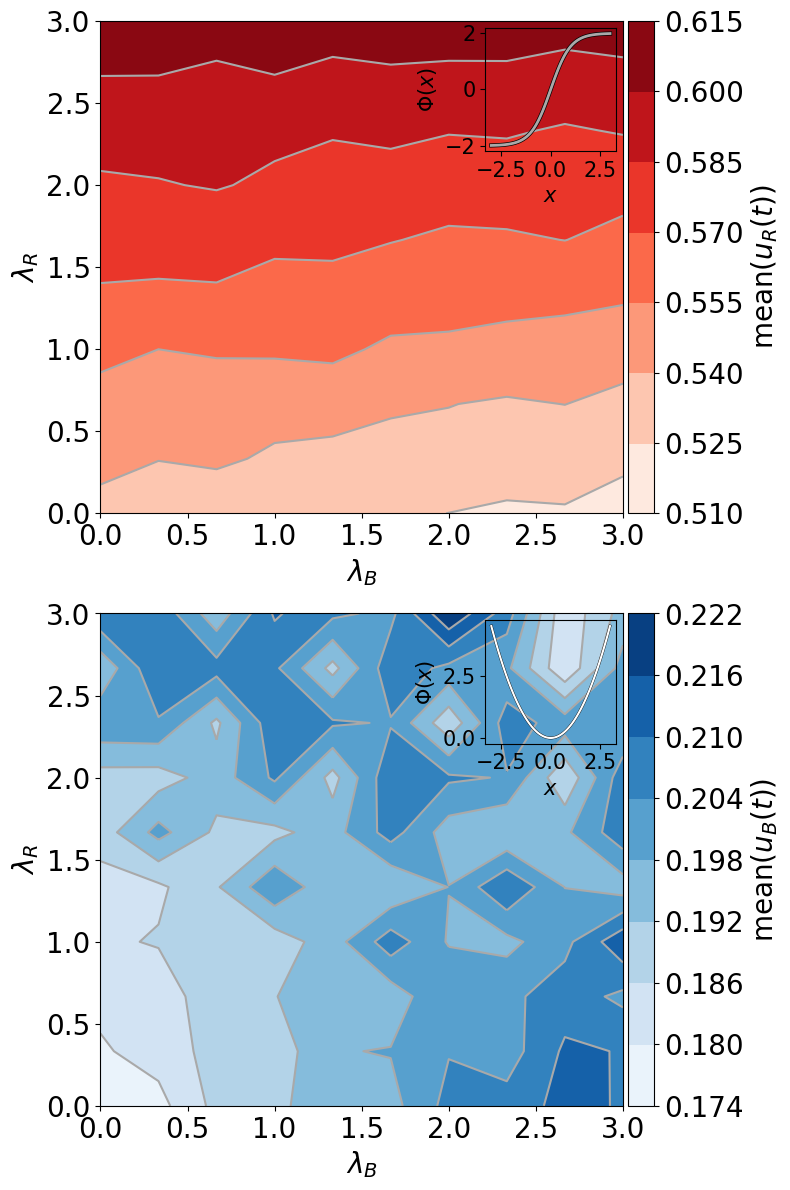}
\caption{Parameter sweep over coupling parameters $\lambda_{\mathrm{R}}, \lambda_{\mathrm{B}}$ with Red final condition $\Phi_{\mathrm{R}}(x) =$ $\tanh(x)$ and Blue final condition $\Phi_{\mathrm{B}}(x) =$ $\frac{1}{2}x^2$. Intensity of color corresponds to mean of control policy.}
\end{figure}

%% file: red_blue_lambda_std_appendix.tex
\begin{figure}[!ht]
    \centering
\includegraphics[width=0.9\linewidth]{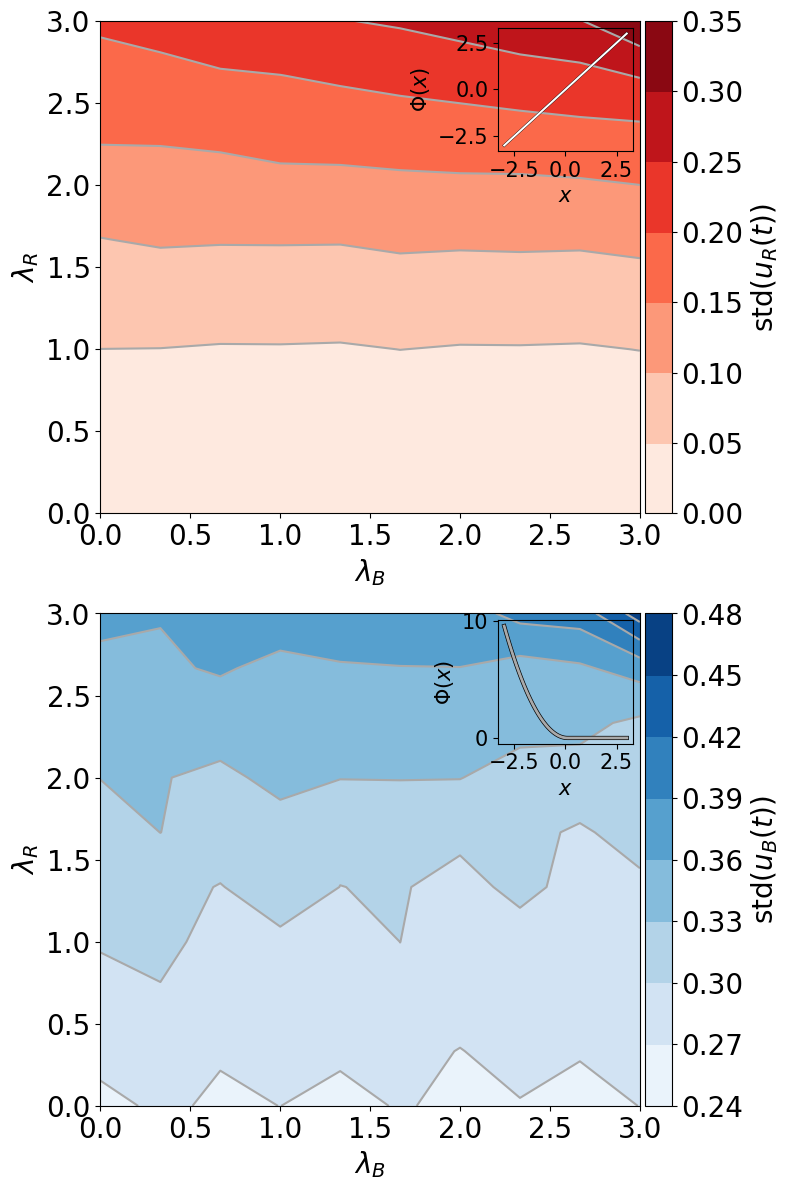}
\caption{Parameter sweep over coupling parameters $\lambda_{\mathrm{R}}, \lambda_{\mathrm{B}}$ with Red final condition $\Phi_{\mathrm{R}}(x) =$ $x$ and Blue final condition $\Phi_{\mathrm{B}}(x) =$ $\frac{1}{2}x^2\Theta(-x)$. Intensity of color corresponds to std of control policy.}
\end{figure}
\begin{figure}[!ht]
    \centering
\includegraphics[width=0.9\linewidth]{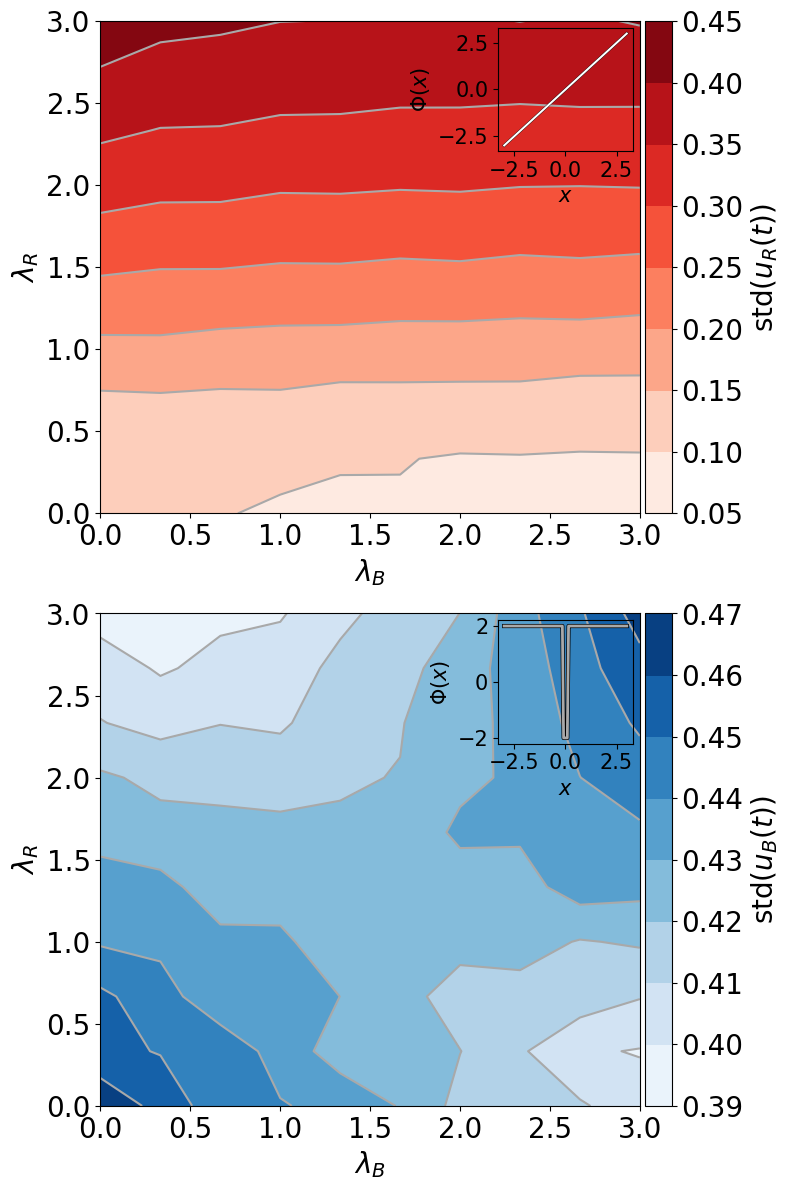}
\caption{Parameter sweep over coupling parameters $\lambda_{\mathrm{R}}, \lambda_{\mathrm{B}}$ with Red final condition $\Phi_{\mathrm{R}}(x) =$ $x$ and Blue final condition $\Phi_{\mathrm{B}}(x) =$ $2[\Theta(|x| - 0.1) - \Theta(0.1 - |x|)]$. Intensity of color corresponds to std of control policy.}
\end{figure}
\clearpage
\begin{figure}[!ht]
    \centering
\includegraphics[width=0.9\linewidth]{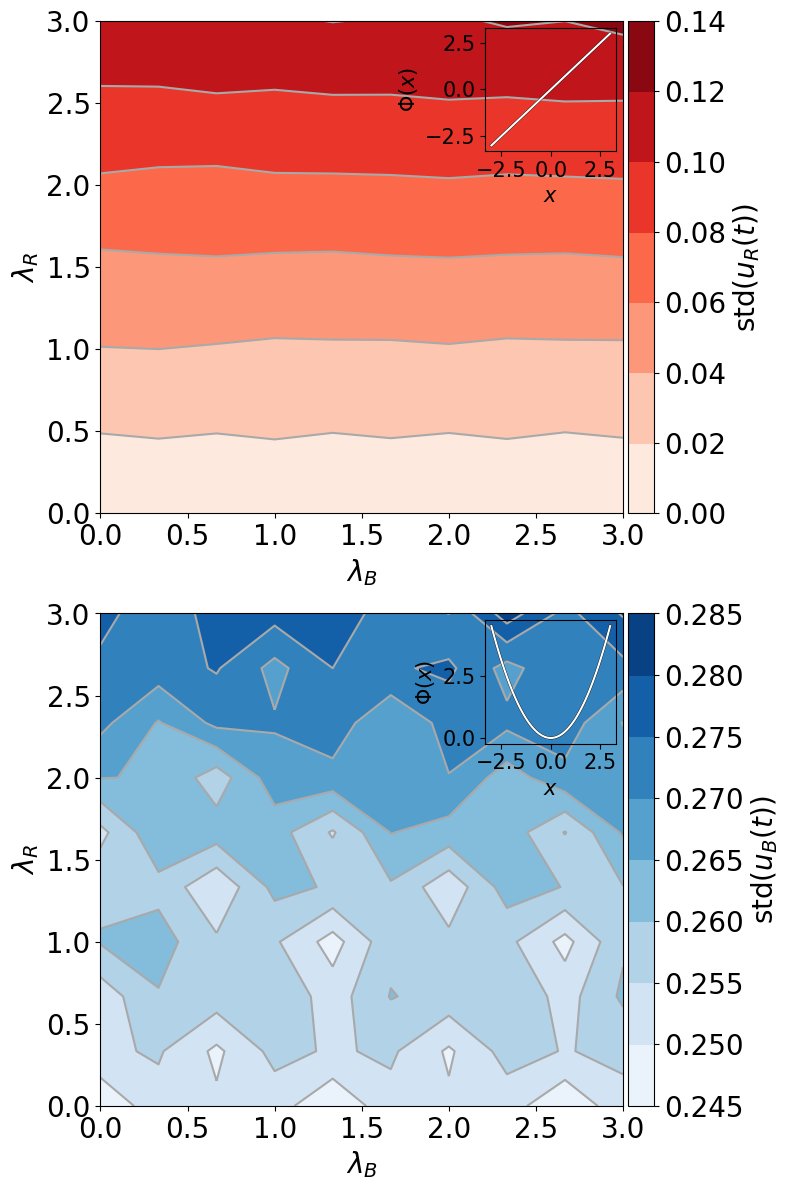}
\caption{Parameter sweep over coupling parameters $\lambda_{\mathrm{R}}, \lambda_{\mathrm{B}}$ with Red final condition $\Phi_{\mathrm{R}}(x) =$ $x$ and Blue final condition $\Phi_{\mathrm{B}}(x) =$ $\frac{1}{2}x^2$. Intensity of color corresponds to std of control policy.}
\end{figure}
\begin{figure}[!ht]
    \centering
\includegraphics[width=0.9\linewidth]{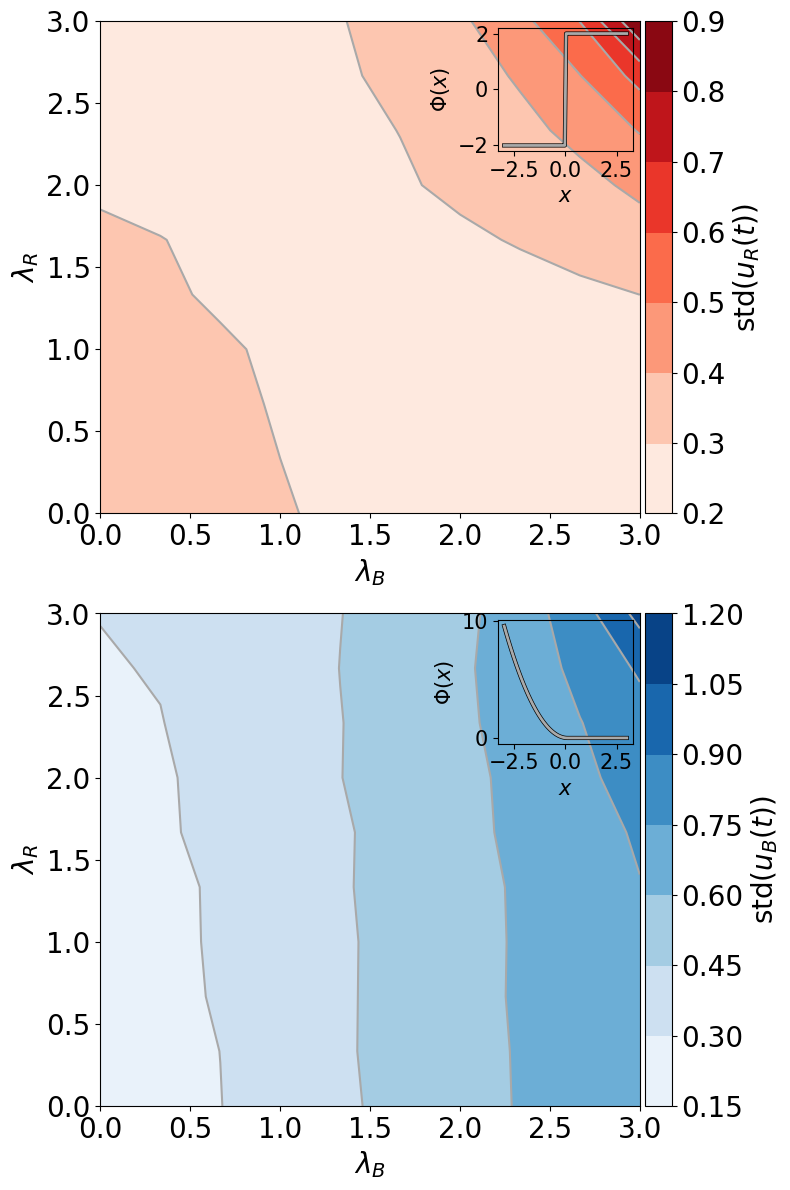}
\caption{Parameter sweep over coupling parameters $\lambda_{\mathrm{R}}, \lambda_{\mathrm{B}}$ with Red final condition $\Phi_{\mathrm{R}}(x) =$ $2[\Theta(x) - \Theta(-x)]$ and Blue final condition $\Phi_{\mathrm{B}}(x) =$ $\frac{1}{2}x^2\Theta(-x)$. Intensity of color corresponds to std of control policy.}
\end{figure}
\clearpage
\begin{figure}[!ht]
    \centering
\includegraphics[width=0.9\linewidth]{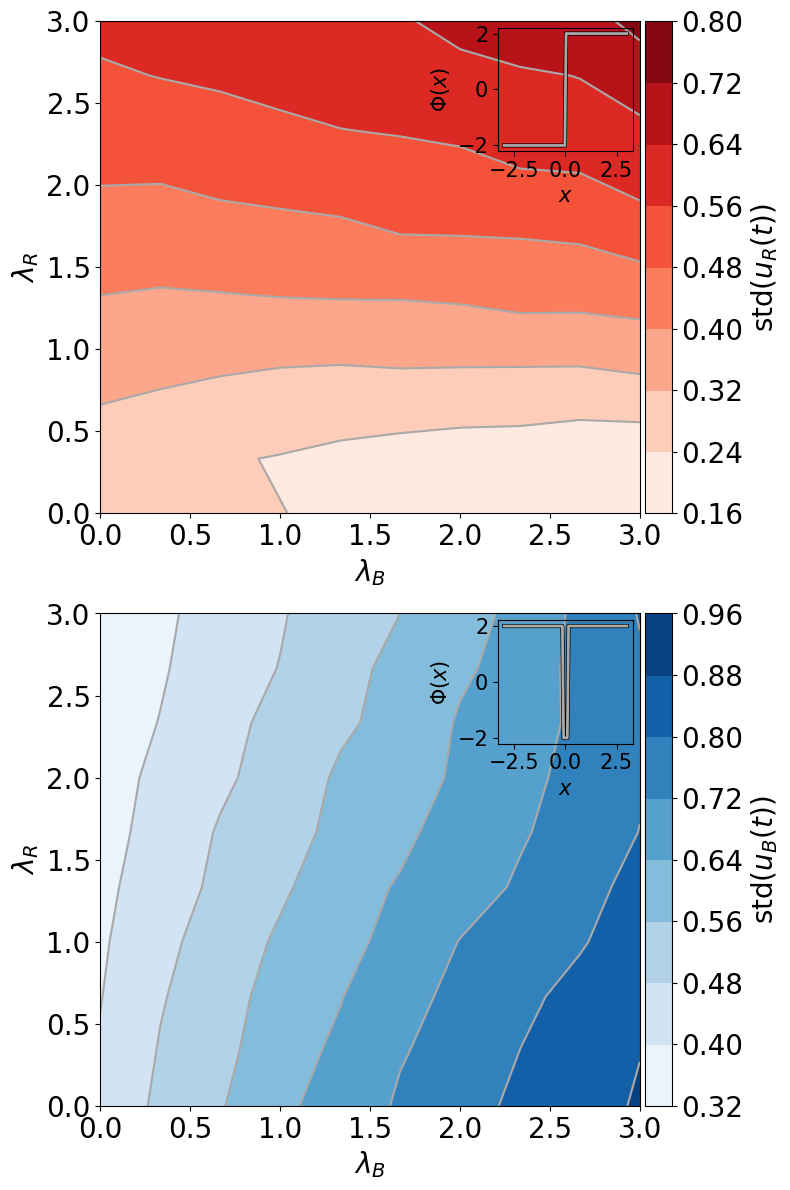}
\caption{Parameter sweep over coupling parameters $\lambda_{\mathrm{R}}, \lambda_{\mathrm{B}}$ with Red final condition $\Phi_{\mathrm{R}}(x) =$ $2[\Theta(x) - \Theta(-x)]$ and Blue final condition $\Phi_{\mathrm{B}}(x) =$ $2[\Theta(|x| - 0.1) - \Theta(0.1 - |x|)]$. Intensity of color corresponds to std of control policy.}
\end{figure}
\begin{figure}[!ht]
    \centering
\includegraphics[width=0.9\linewidth]{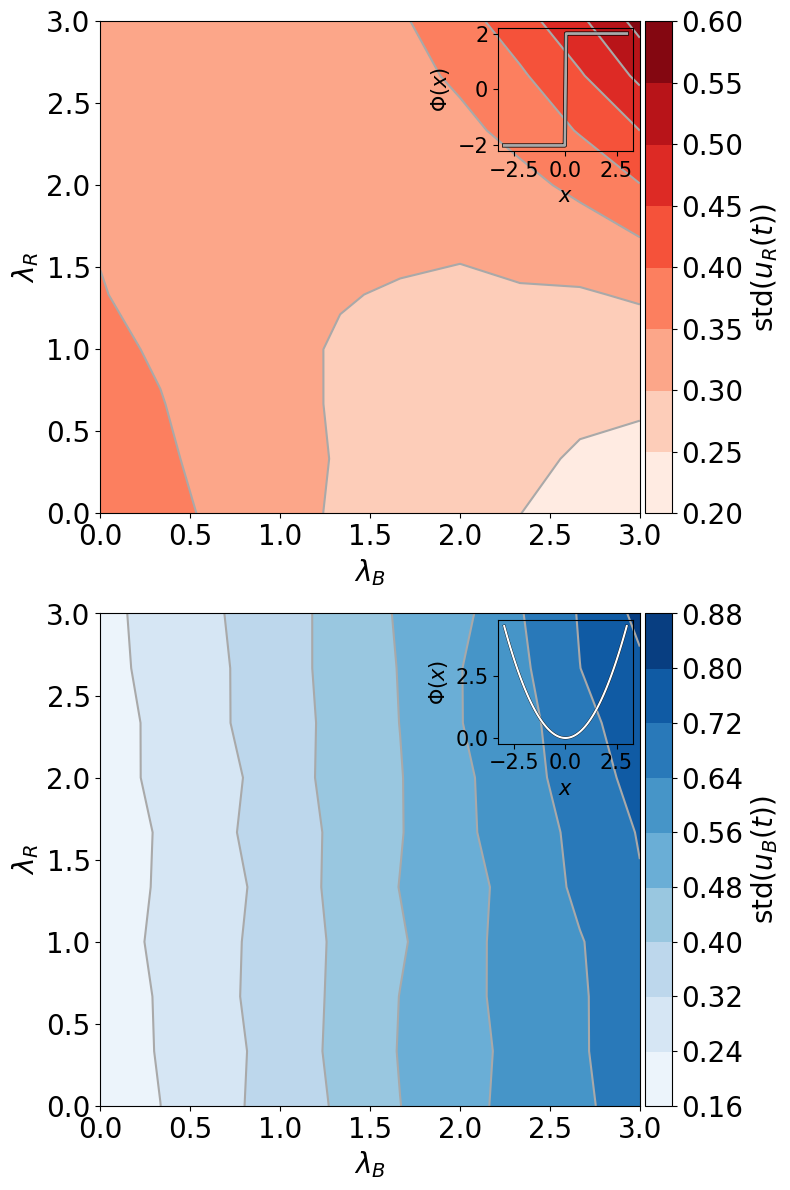}
\caption{Parameter sweep over coupling parameters $\lambda_{\mathrm{R}}, \lambda_{\mathrm{B}}$ with Red final condition $\Phi_{\mathrm{R}}(x) =$ $2[\Theta(x) - \Theta(-x)]$ and Blue final condition $\Phi_{\mathrm{B}}(x) =$ $\frac{1}{2}x^2$. Intensity of color corresponds to std of control policy.}
\end{figure}
\clearpage
\begin{figure}[!ht]
    \centering
\includegraphics[width=0.9\linewidth]{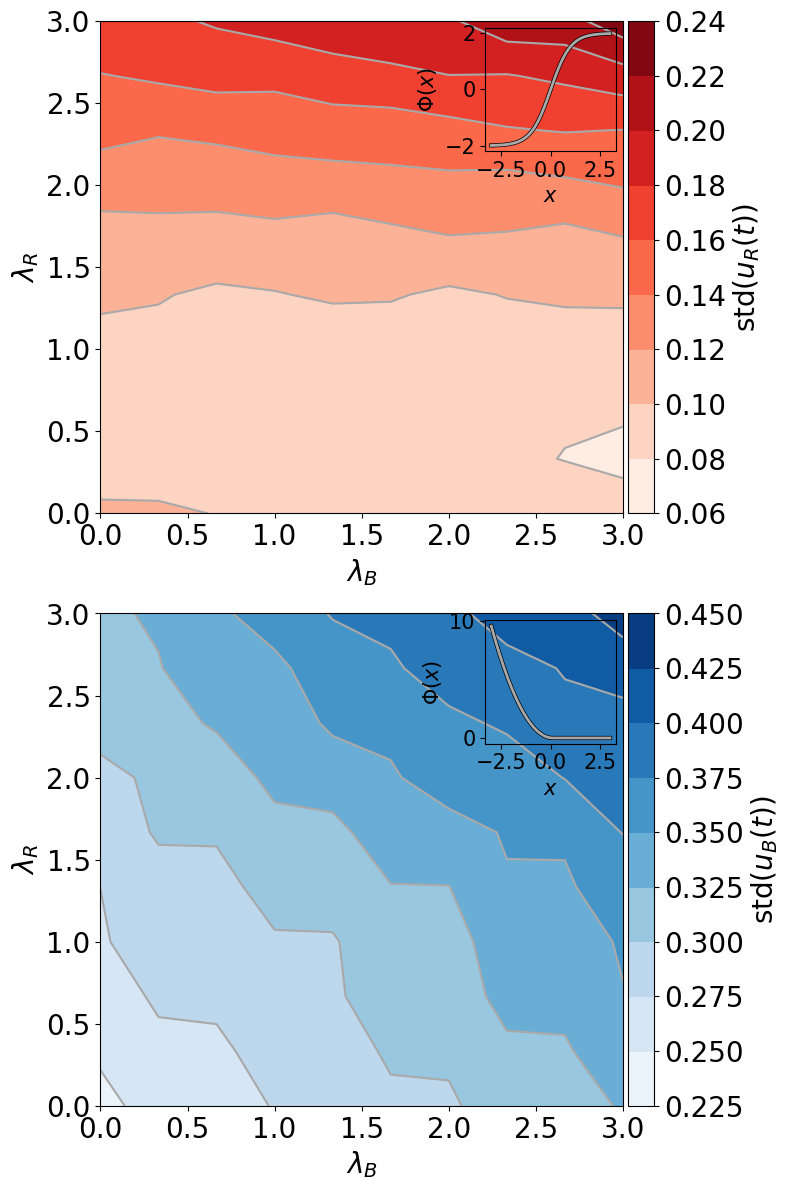}
\caption{Parameter sweep over coupling parameters $\lambda_{\mathrm{R}}, \lambda_{\mathrm{B}}$ with Red final condition $\Phi_{\mathrm{R}}(x) =$ $\tanh(x)$ and Blue final condition $\Phi_{\mathrm{B}}(x) =$ $\frac{1}{2}x^2\Theta(-x)$. Intensity of color corresponds to std of control policy.}
\end{figure}
\begin{figure}[!ht]
    \centering
\includegraphics[width=0.9\linewidth]{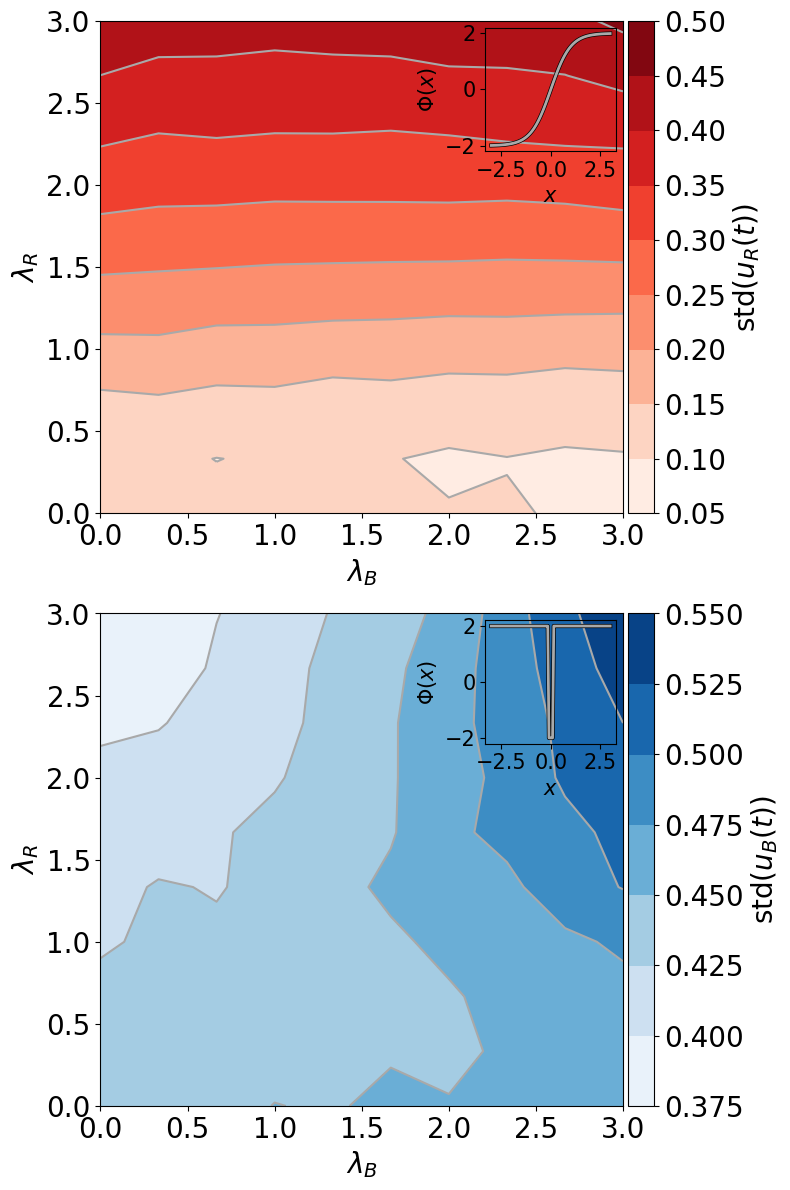}
\caption{Parameter sweep over coupling parameters $\lambda_{\mathrm{R}}, \lambda_{\mathrm{B}}$ with Red final condition $\Phi_{\mathrm{R}}(x) =$ $\tanh(x)$ and Blue final condition $\Phi_{\mathrm{B}}(x) =$ $2[\Theta(|x| - 0.1) - \Theta(0.1 - |x|)]$. Intensity of color corresponds to std of control policy.}
\end{figure}
\clearpage
\begin{figure}[!ht]
    \centering
\includegraphics[width=0.9\linewidth]{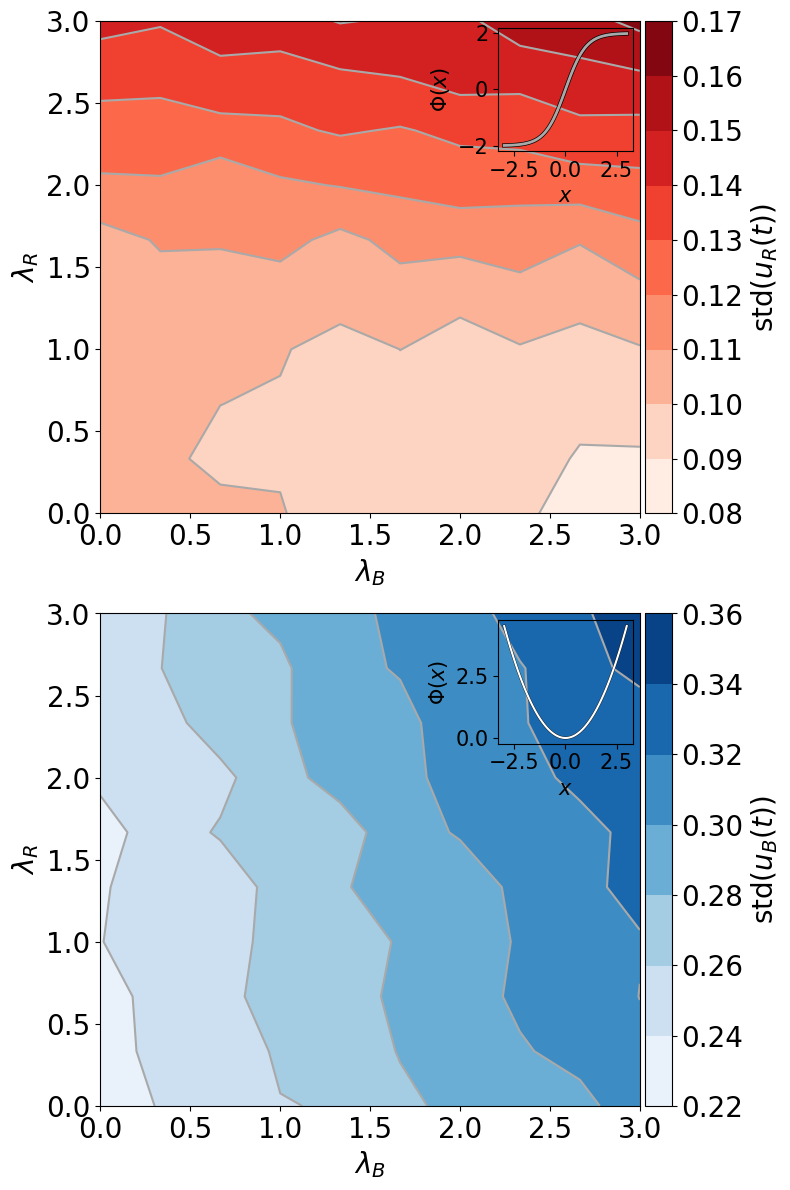}
\caption{Parameter sweep over coupling parameters $\lambda_{\mathrm{R}}, \lambda_{\mathrm{B}}$ with Red final condition $\Phi_{\mathrm{R}}(x) =$ $\tanh(x)$ and Blue final condition $\Phi_{\mathrm{B}}(x) =$ $\frac{1}{2}x^2$. Intensity of color corresponds to std of control policy.}
\end{figure}